%% file: arxiv_main.tex
\documentclass[twocolumn,11pt]{article}

\usepackage{helvet}
\usepackage{graphicx,caption,subcaption}	% Including figure files
\usepackage{amsmath}	% Advanced maths commands
\usepackage{amssymb}	% Extra maths symbols
\usepackage{textcomp}
\usepackage{gensymb} %math symbols, e.g. \degree
\usepackage{color}
\usepackage{enumitem}
\usepackage{multirow}
\usepackage{verbatim} %for comment command
\usepackage{authblk} %for affiliations
\usepackage[switch]{lineno}
\usepackage{hyperref}
\usepackage[margin=1in]{geometry}
\newcommand{\aj}{Astron. J.}

\newcommand{\apj}{Astrophys. J.}
\newcommand{\apjl}{Astrophys. J., Letters}
\newcommand{\apjs}{Astrophys. J., Suppl. Ser.}

\newcommand{\apss}{Astrophysics and Space Science}
\newcommand{\aap}{Astron. Astrophys.}

\newcommand{\icarus}{Icarus}

\newcommand{\mnras}{Mon. Not. R. Astron. Soc.}

\newcommand{\pasp}{Publ. Astron. Soc. Pacific}

\newcommand{\pnas}{Proc. Natl. Acad. Sci.}

\newcommand{\nat}{Nature}

\newcommand{\procspie}{Proc. SPIE}

\newcommand{\Ntarget}{TOI-270}

\newcommand{\Nplanetb}{TOI-270\,b}
\newcommand{\Nplanetc}{TOI-270\,c}
\newcommand{\Nplanetd}{TOI-270\,d}

\newcommand{\Ntransits}{42}

\newcommand{\RA}{\mbox{$04^{\mathrm{h}} 33^{\mathrm{m}} 39.72^{\mathrm{s}} $}} %04 33 39.7202257810  Gaia DR2 RAJ2000
\newcommand{\Dec}{\mbox{$-51{\degree} 57' 22.44''$}} %-51 57 22.436253190 Gaia DR2 DECJ2000
\newcommand{\Longecl}{\mbox{$02^{\mathrm{h}} 52^{\mathrm{m}} 35.24^{\mathrm{s}}$}}
\newcommand{\Latecl}{\mbox{$-71{\degree} 53' 49.29''$}}
\newcommand{\toiid}{270}
\newcommand{\ticid}{259377017}
\newcommand{\twomassid}{\mbox{J04333970--5157222}} %2MASS
\newcommand{\gaiaid}{\mbox{4781196115469953024}} %GAIA DR2 
\newcommand{\propRA}{\mbox{$82.944\pm0.050$}} %GAIA DR2
\newcommand{\propDec}{\mbox{$-269.755\pm0.051$}} %GAIA DR2
\newcommand{\TESSmag}{$10.416$}
\newcommand{\Vmag}{$12.62 \pm 0.03$} %UCAC4
\newcommand{\gmag}{$13.391 \pm 0.02$} %UCAC4
\newcommand{\rmag}{$12.011 \pm 0.02$} %UCAC4
\newcommand{\imag}{$10.910 \pm 0.059$} %UCAC4
\newcommand{\GAIAmag}{$11.6306$} %GAIA DR2
\newcommand{\GAIAbpmag}{$12.87021$} %GAIA DR2
\newcommand{\GAIArpmag}{$10.54313$} %GAIA DR2
\newcommand{\Jmag}{$9.099 \pm 0.032$} %2MASS
\newcommand{\Hmag}{$8.531 \pm 0.073$} %2MASS
\newcommand{\Kmag}{$8.251 \pm 0.029$} %2MASS

\newcommand{\starmetallicity}{$-0.17 \pm 0.1$}

\newcommand{\parallax}{\mbox{$44.538\pm0.043$}}% GAIA DR2 + Stassun&Torres 2018
\newcommand{\distancely}{\mbox{$73.231 \pm 0.070$}}
\newcommand{\distancepc}{\mbox{$22.453 \pm 0.021$}}

\newcommand{\Kmagabs}{$6.490 \pm 0.029$}
\newcommand{\starmassBenedictsixteen}{$0.40 \pm 0.02$}
\newcommand{\starradiusBenedictsixteenBoyajiantwelve}{$0.38 \pm 0.02$}
\newcommand{\starmassMannnineteen}{$0.36 \pm 0.02$}

\newcommand{\starradiusMannfifteen}{$0.37 \pm 0.02$}
\newcommand{\starBCK}{$2.6 \pm 0.1$}
\newcommand{\starbolometricmagnitude}{$9.1 \pm 0.1$}
\newcommand{\starbolometricluminosity}{$0.017 \pm 0.002$}
\newcommand{\starteff}{$3386_{-131}^{+137}$}
\newcommand{\startype}{$\mathrm{M}3.0\pm0.5\mathrm{V}$}

\newcommand{\kepler}{{\it Kepler}}
\newcommand{\tess}{{\it TESS}}
\newcommand{\gaia}{{\it Gaia}}
\newcommand{\twomass}{2MASS}
\newcommand{\jwst}{{\it JWST}}
\newcommand{\spitzer}{{\it Spitzer}}
\newcommand{\harps}{{\it HARPS}}
\newcommand{\espresso}{{\it ESPRESSO}}

\newcommand{\mstar}{\mbox{$M_\star$}}
\newcommand{\rstar}{\mbox{$R_\star$}}
\newcommand{\teff}{$T_{\rm eff}$}

\newcommand{\msun}{\mbox{M$_{\odot}$}}
\newcommand{\rsun}{\mbox{R$_{\odot}$}}
\newcommand{\rearth}{R$_{\oplus}$}
\newcommand{\mearth}{M$_{\oplus}$}

\newcommand{\masy}{mas\,yr$^{-1}$}
\newcommand{\gccc}{g\,cm$^{-3}$}

\newcommand{\bperiodshort}{$3.36$} %none = 3.360062_{-0.000095}^{+0.000088}
\newcommand{\cperiodshort}{$5.66$} %none = 5.660172\pm0.000040
\newcommand{\dperiodshort}{$11.38$} %none = 11.38028\pm0.00017

\newcommand{\bRcompanionRearth}{$1.247_{-0.083}^{+0.089}$} %$R_\mathrm{b}$ ($\mathrm{R_{\oplus}}$) = 1.247_{-0.083}^{+0.089}
\newcommand{\bTeq}{$528_{-32}^{+56}$} %$T_\mathrm{eq;b}$ (K) = 528_{-32}^{+56}
\newcommand{\cRcompanionRearth}{$2.42\pm0.13$} %$R_\mathrm{c}$ ($\mathrm{R_{\oplus}}$) = 2.42\pm0.13
\newcommand{\cTeq}{$424_{-19}^{+20}$} %$T_\mathrm{eq;c}$ (K) = 424_{-19}^{+20.}
\newcommand{\dRcompanionRearth}{$2.13\pm0.12$} %$R_\mathrm{d}$ ($\mathrm{R_{\oplus}}$) = 2.13\pm0.12
\newcommand{\dTeq}{$340\pm14$} %$T_\mathrm{eq;d}$ (K) = 340.\pm14
\newcommand{\meanhostdensity}{$10.5_{-2.2}^{+1.4}$} %$rho_\mathrm{\star; mean}$ (cgs) = 10.5_{-2.2}^{+1.4}

\newcommand{\bMcompanionMearth}{$1.9^{+1.5}_{-0.7}$} 
\newcommand{\cMcompanionMearth}{$6.6^{+5.2}_{-2.8}$} 
\newcommand{\dMcompanionMearth}{$5.4^{+4.0}_{-2.1}$}

\date{}
\title{\vspace{-0.5in}A super-Earth and two sub-Neptunes transiting the nearby and quiet M~dwarf TOI-270}

\author[*,1,2]{Maximilian N.\ G{\"u}nther}

\author[3,4]{Francisco J.\ Pozuelos}  
\author[5,6]{Jason A.\ Dittmann}    
\author[1,7]{Diana Dragomir} 
\author[8]{Stephen R.\ Kane}  
\author[1,9]{Tansu Daylan}        
\author[10]{Adina D.\ Feinstein}
\author[1,2]{Chelsea Huang}     
\author[11]{Timothy D.\ Morton}          

\author[3]{Andrea Bonfanti}       
\author[18]{L. G. Bouma}           
\author[1,2]{Jennifer Burt}       
\author[12]{Karen A.\ Collins}      
\author[13]{Jack J. Lissauer}      
\author[1]{Elisabeth Matthews}     
\author[10, 16]{Benjamin T.\ Montet} 
\author[15,16]{Andrew Vanderburg}    
\author[17,6]{Songhu Wang}          
\author[12]{Jennifer G. Winters}  

\author[1]{George R.\ Ricker}       
\author[1]{Roland K.\ Vanderspek}    
\author[12]{David W.\ Latham}       
\author[1,5,14]{Sara Seager}         
\author[18]{Joshua N.\ Winn}     
\author[13]{Jon M.\ Jenkins}     

\author[19]{James D.\ Armstrong}   
\author[4,20]{Khalid Barkaoui}     
\author[21]{Natalie Batalha}       
\author[10]{Jacob L.\ Bean}         
\author[23]{Douglas A. Caldwell}  
\author[23]{David R. Ciardi}     
\author[24]{Kevin I.\ Collins}    
\author[1]{Ian Crossfield}        
\author[1]{Michael Fausnaugh} 	   
\author[1]{Gabor Furesz} 	      
\author[25]{Tianjun Gan}           
\author[4]{Micha\"el Gillon}      
\author[1]{Natalia Guerrero}      
\author[26]{Keith Horne}          
\author[13]{Steve B.\ Howell}      
\author[27]{Michael Ireland}        
\author[28]{Giovanni Isopi}      
\author[3]{Emmanu\"el Jehin}      
\author[29]{John F.\ Kielkopf}    
\author[30]{Sebastien Lepine}    
\author[28]{Franco Mallia}        
\author[13]{Rachel A.\ Matson}  
\author[31]{Gordon Myers}         
\author[32,33]{Enric Palle}       
\author[12]{Samuel N.\ Quinn}     
\author[1]{Howard M. Relles}      
\author[34]{B\'arbara Rojas-Ayala}  
\author[35]{Joshua Schlieder}     
\author[36]{Ramotholo Sefako}  
\author[1]{Avi Shporer}         
\author[37,38]{Juan C. Su\'arez}   
\author[39]{Thiam-Guan Tan}       
\author[13]{Eric B. Ting} 	       
\author[22]{Joseph D. Twicken} 	   
\author[40]{Ian A. Waite}

\affil[1]{Department of Physics, and Kavli Institute for Astrophysics and Space Research, MIT, Cambridge, MA 02139, USA}
\affil[2]{Juan Carlos Torres Fellow}
\affil[3]{Space Sciences, Technologies and Astrophysics Research (STAR) Institute, Universit{\'e} de Li{\`e}ge, 19C All{\'e}e du 6 Ao\t{u}t, 4000 Li{\`e}ge, Belgium}
\affil[4]{Astrobiology Research Unit, Universit{\'e} de Li{\`e}ge, 19C Allée du 6 Ao\t{u}t, 4000 Li{\`e}ge, Belgium}
\affil[5]{Department of Earth, Atmospheric, and Planetary Sciences, MIT, Cambridge, MA 02139, USA}
\affil[6]{51 Pegasi b Postdoctoral Fellow}
\affil[7]{NASA Hubble Fellow}
\affil[8]{Department of Earth and Planetary Sciences, University of California, Riverside, CA 92521, USA}
\affil[9]{Kavli Fellow}
\affil[10]{Department of Astronomy \& Astrophysics, University of Chicago, 5640 S.\ Ellis Avenue, Chicago, IL 60637, USA}
\affil[11]{Department of Astronomy, University of Florida, 211 Bryant Space Science Center, Gainesville, FL, 32611, USA}
\affil[12]{Center for Astrophysics | Harvard \& Smithsonian, 60 Garden Street, Cambridge, MA 02138}
\affil[13]{NASA Ames Research Center, Moffett Field, CA, 94035, USA}
\affil[14]{Department of Aeronautical and Astronautical Engineering, MIT, Cambridge, MA 02139, USA}
\affil[15]{Department of Astronomy, The University of Texas at Austin, Austin, TX 78712, USA}
\affil[16]{NASA Sagan Fellow}
\affil[17]{Department of Astronomy, Yale University, New Haven, CT 06511, USA}
\affil[18]{Department of Astrophysical Sciences, Princeton University, 4 Ivy Lane, Princeton, NJ 08544, USA}
\affil[19]{University of Hawaii Institute for Astronomy, 34 Ohia Ku Street, Pukalani, HI 96753}
\affil[20]{Oukaimeden Observatory, High Energy Physics and Astrophysics Laboratory, Cadi Ayyad University, Marrakech, Morocco}
\affil[21]{Department of Astronomy and Astrophysics, University of California, Santa Cruz, CA 95064, USA}
\affil[22]{SETI Institute/NASA Ames Research Center, Moffett Field, CA 94035, USA}
\affil[23]{NASA Exoplanet Science Institute, Caltech/IPAC-NExScI, 1200 East California Boulevard, Pasadena, CA 91125, USA}
\affil[24]{Department of Physics and Astronomy, Vanderbilt University, Nashville, TN 37235, USA}
\affil[25]{Physics Department and Tsinghua Centre for Astrophysics, Tsinghua University, Beijing 100084, China}
\affil[26]{SUPA Physics \& Astronomy, University of St~Andrews,  North Haugh, St~Andrews, KY16~9SS, Scotland, UK}
\affil[27]{Research School of Astronomy and Astrophysics, Australian National University, Canberra, ACT 2611, Australia}
\affil[28]{Campo Catino Astronomical Observatory, Regione Lazio, Guarcino (FR), 03010 Italy}
\affil[29]{Department of Physics and Astronomy, University of Louisville, Louisville, KY 40292, USA}
\affil[30]{Department of Physics and Astronomy, Georgia State University, 25 Park Pl. NE, Atlanta, GA 30340, USA}
\affil[31]{AAVSO, 5 Inverness Way, Hillsborough, CA 94010, USA}
\affil[32]{Instituto de Astrof\'\i sica de Canarias (IAC), 38205 La Laguna, Tenerife, Spain}
\affil[33]{Departamento de Astrof\'\i sica, Universidad de La Laguna (ULL), 38206 La Laguna, Tenerife, Spain}
\affil[34]{Departamento de Ciencias F\'isicas, Universidad Andr\'es Bello, Fern\'andez Concha 700, Las Condes, Santiago, Chile}
\affil[35]{NASA Goddard Space Flight Center, Greenbelt, MD, USA}
\affil[36]{South African Astronomical Observatory, PO Box 9, Observatory, 7935, South Africa}
\affil[37]{Dpt. F\'isica Te\'orica y del Cosmos, Universidad de Granada, Campus de Fuentenueva s/n, 18071, Granada, Spain }
\affil[38]{Instituto de Astrof\'isica de Andaluc\'ia (CSIC), Glorieta de la Astronom\'ia s/n, 18008, Granada, Spain}
\affil[39]{Perth Exoplanet Survey Telescope, Perth, Western Australia}
\affil[40]{Centre for Astrophysics, University of Southern Queensland, Toowoomba, QLD, 4350, Australia\vspace*{11pt}}
\affil[*]{Corresponding author (\url{maxgue@mit.edu})}

\begin{document}

\maketitle
\clearpage

\include{arxiv_letter}
\include{arxiv_methods}

\include{arxiv_supplement}

\end{document}

%% file: arxiv_letter.tex
% Intro / Abstract
{\bf
One of the primary goals of exoplanetary science is to detect small, temperate planets passing (transiting) in front of bright and quiet host stars. This enables the characterisation of planets' sizes, orbits, bulk compositions, atmospheres and formation histories. These studies are further favoured by small and cool M~dwarf hosts.
Here, we report the Transiting Exoplanet Survey Satellite discovery of three small planets transiting one of the nearest and brightest M~dwarf hosts to date, {\Ntarget} (TIC~{\ticid}; K-mag~8.3; 22.5~parsec).
The M3V-type star is transited by the super-Earth-sized {\Nplanetb} ({\bRcompanionRearth}~{\rearth}) and the sub-Neptune-sized 
{\Nplanetc} ({\cRcompanionRearth}~{\rearth}) and {\Nplanetd} ({\dRcompanionRearth}~{\rearth}).
The planets orbit close to a mean-motion resonant chain, with periods ({\bperiodshort}, {\cperiodshort}, and {\dperiodshort} days) near ratios of small integers (\mbox{$5:3$} and \mbox{$2:1$}).
{\Ntarget} is a prime target for future studies since: 
1) its near-resonance allows detecting transit timing variations for precise mass measurements and dynamical studies; 
2) its brightness enables independent radial velocity mass measurements; 
3) the outer planets are ideal for atmospheric characterisation via transmission spectroscopy;
and 4) the quiet star enables future searches for habitable zone planets.
Altogether, very few systems with small, temperate exoplanets are as suitable for such complementary and detailed characterisation as {\Ntarget}.
}

\begin{figure*}[!htbp]
    \centering
    \begin{subfigure}[b]{0.8\textwidth}
        \centering
        \includegraphics[width=\columnwidth]{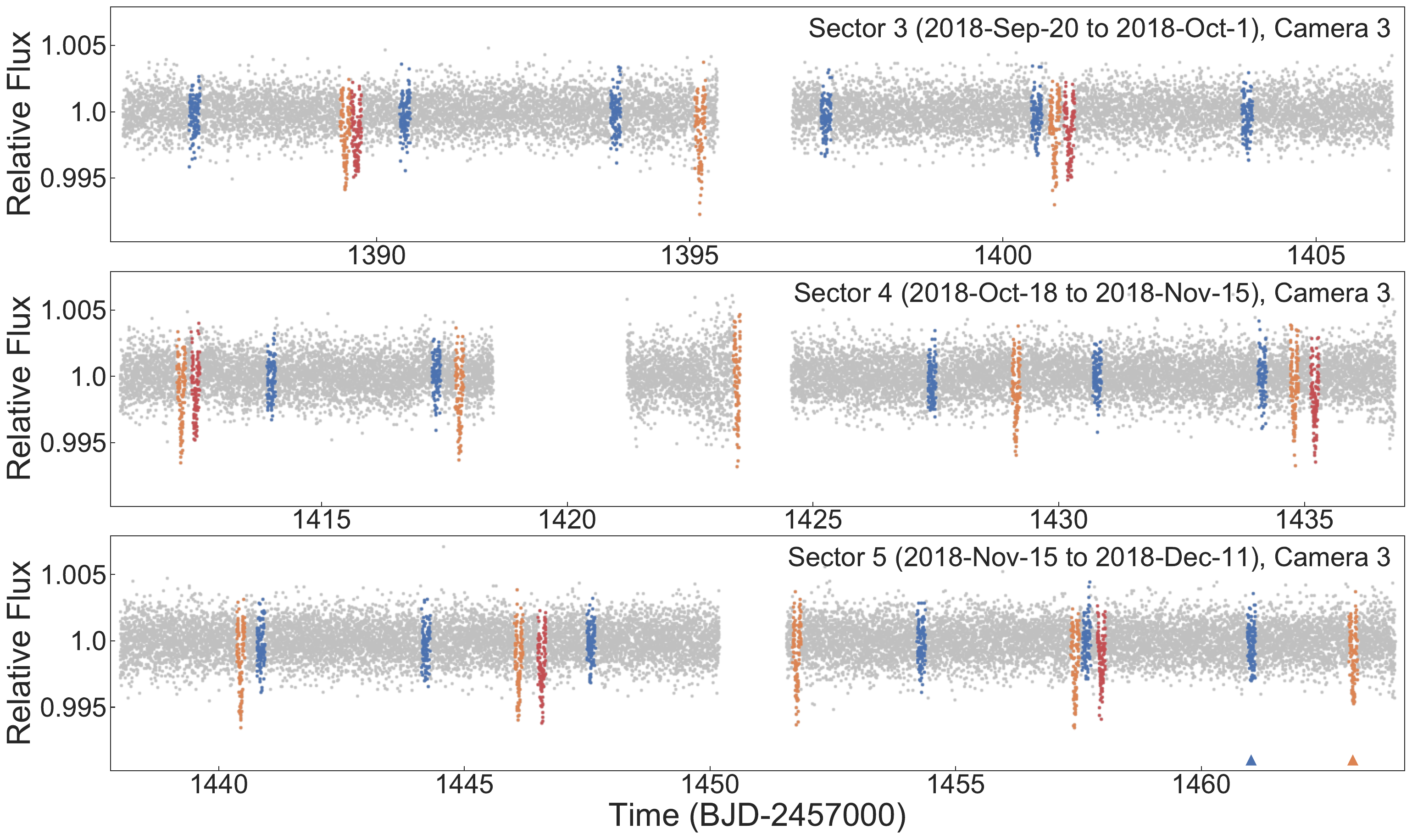}
    \end{subfigure}
    \begin{subfigure}[b]{0.41\textwidth}
        \centering
        \includegraphics[width=\columnwidth]{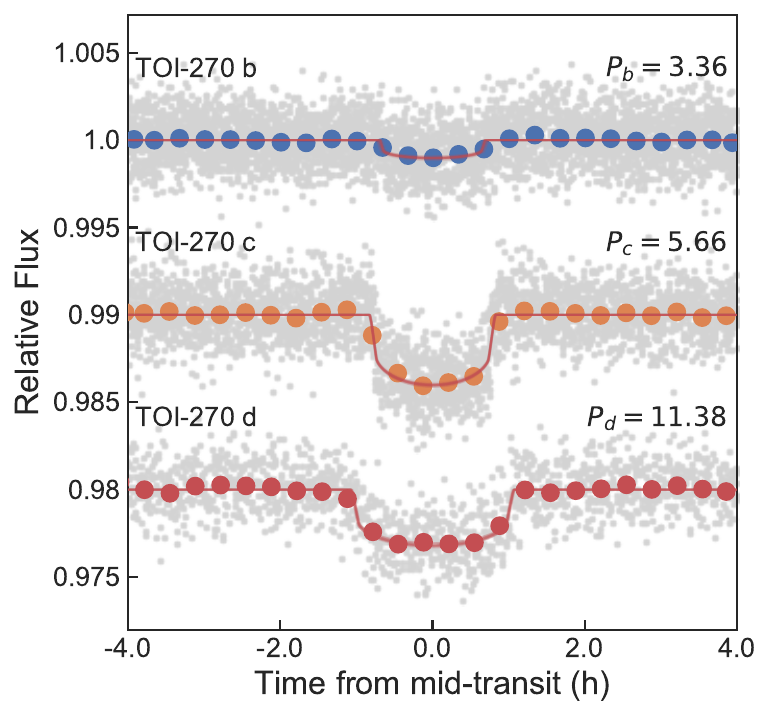}
    \end{subfigure}
    \begin{subfigure}[b]{0.4\textwidth}
        \centering
        \includegraphics[width=\columnwidth]{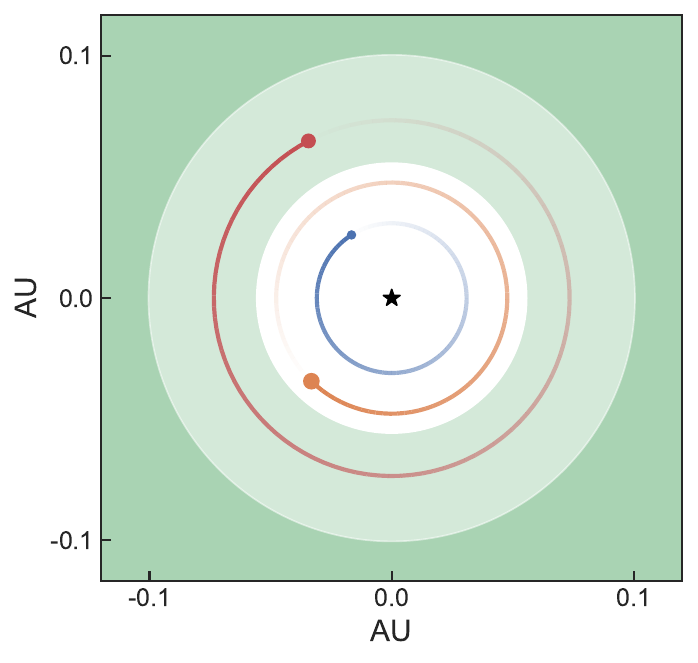}
    \end{subfigure}
    \caption{
    The discovery data and orbits of the super-Earth and two sub-Neptunes transiting {\Ntarget}.
    \textit{Top:} the full {\tess} discovery lightcurve of Sectors~3, 4 and 5 (grey points), with transits of planets b (blue), c (orange) and d (red) marked in colour. 
    \textit{Lower left:} {\tess} lightcurves phase-folded onto the best-fit periods for all three planets. Grey points show the individual 2~min. cadence observations, coloured circles show the data binned in phase with 15~min. spacing. Red lines show 20 lightcurve models generated from randomly drawn posterior samples by the \texttt{allesfitter} analysis (\cite{allesfitter}; see Methods).
    \textit{Lower right:} a top-down view of the system. The dark-green area shows the optimistic habitable zone according to \cite{Kopparapu2014}, spanning 0.10~AU to 0.28~AU. The light-green area shows the orbital distance at which the equilibrium temperature of a planet is between the survival temperature for extremophiles ($\sim$395~K) \cite{Takai2008} and the freezing point of water (273.15~K), spanning 0.06~AU to 0.12~AU. Note that the equilibrium temperature can differ from the surface temperature.
    }
    \label{fig:TESS_lightcurve}
\end{figure*}

\begin{figure*}[!htbp]
    \centering
    \includegraphics[width=0.8\textwidth]{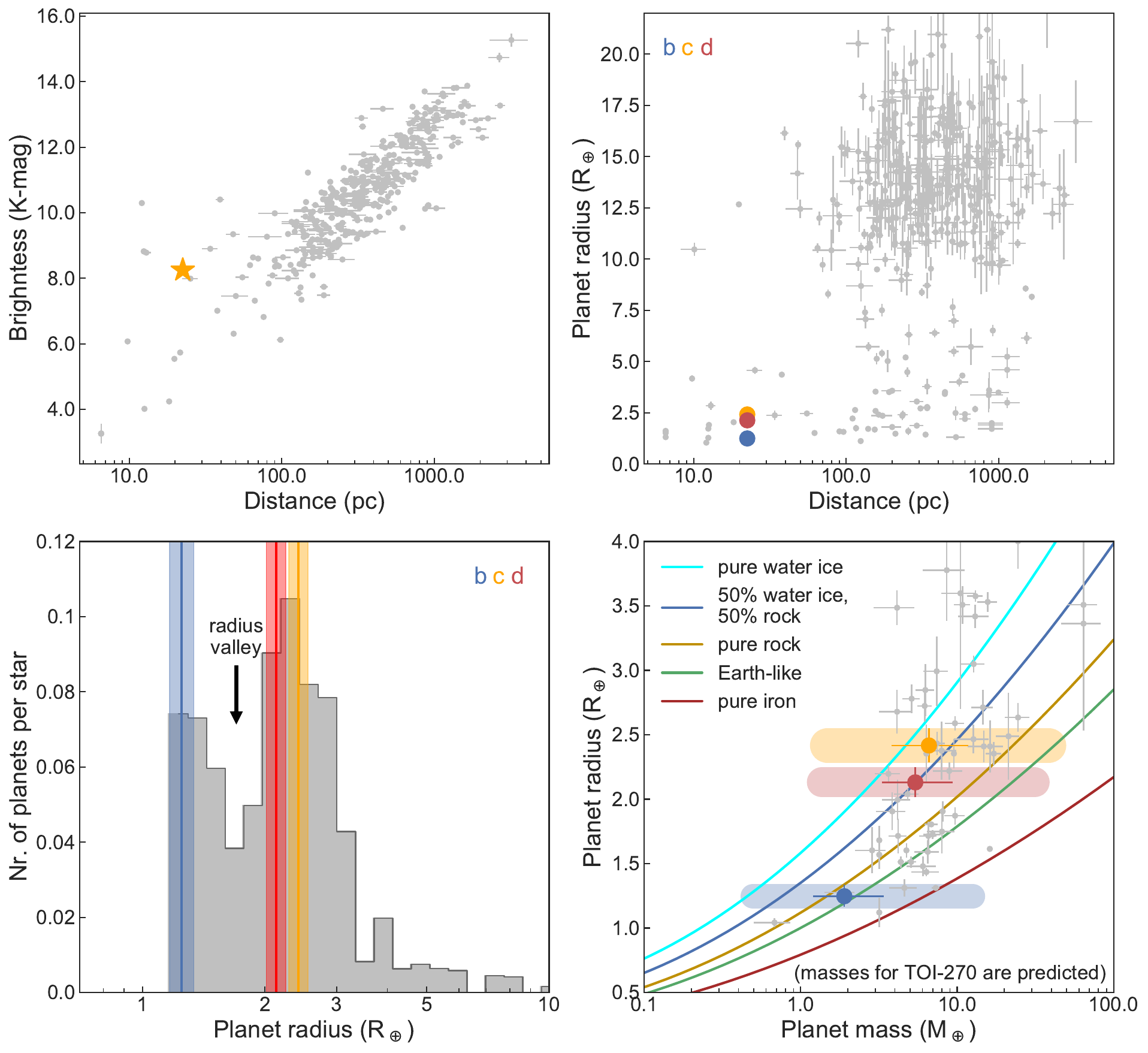}
    \caption{
            {\Ntarget} in the context of known exoplanets. 
            \textit{Top left:} the brightness (as {\twomass} K-band magnitude) versus the distance to Earth (in parsec) of the {\Ntarget} host star (orange star symbol), compared with known exoplanet hosts (grey circles with 68.3\% credible intervals).
            \textit{Top right:} the planet radii versus the system's distance to Earth, shown for {\Ntarget}~b, c and d (blue, orange and red circles, respectively) compared with known exoplanets (grey circles with 68.3\% credible intervals).
            \textit{Bottom left:} a histogram of the number of planets per star (for orbital periods $<100$\,days) over planet radius, as reported by \cite{Fulton2018}. The radii of {\Ntarget}~b, c and d are marked for comparison (blue, orange, and red lines; coloured bands showing 68.3\% credible intervals). The radius valley appears around 1.7--2.0\,{\rearth} separating two populations of planets, rocky super-Earths and gas-dominated sub-Neptunes (e.g. \cite{Owen2013}). 
            \textit{Bottom right:} mass-radius-diagram indicating the potential bulk compositions of the three {\Ntarget} planets and known exoplanets. The {\Ntarget} masses are predicted from the relations of \cite{Chen2017}, with the 68.3\% and 99.7\% credible intervals of the predictions shown as lines and bands, respectively. Overplotted are theoretical bulk composition curves from \cite{Fortney2007} (water/ice: $\mathrm{H_2O}$; rock: $\mathrm{Mg2SiO4}$; iron: $\mathrm{Fe}$; Earth-like: 67\% rock / 33\% iron), and a comparison with known exoplanets (grey circles with 68.3\% credible intervals).
            In all panels, the data on known exoplanets is from \url{https://exoplanetarchive.ipac.caltech.edu/}, online 2019 March 04. Only known transiting exoplanets with values measured to better than a 30\% relative error are shown.
            These four views highlight how {\Ntarget} occupies an exciting parameter space for future studies, with the diversity of the system providing an interesting case study for planet formation and photoevaporation.
            }
    \label{fig:TOI-270_in_context}
\end{figure*}

%Intro
The super-Earth-sized and two sub-Neptune-sized planets transiting {\Ntarget} were detected by the Transiting Exoplanet Survey Satellite ({\tess}) mission in Sectors 3--5 (Fig.~\ref{fig:TESS_lightcurve}), and followed up with ground-based multi-wavelength photometry, reconnaissance spectroscopy, and high resolution imaging.
Following an extensive vetting protocol including these observations and archival/catalogue data, we validate the transit signals to be of planetary origin and the host to be a single {\startype} star (see Methods). 
With a distance of only 22.5~parsec, {\Ntarget} is one of the closest transiting exoplanet hosts to Earth (Fig~\ref{fig:TOI-270_in_context}).
We find a stellar mass of {\starmassBenedictsixteen}\,{\msun}, radius of {\starradiusBenedictsixteenBoyajiantwelve}\,{\rsun}, effective temperature of {\starteff}\,K, and metallicity of {\starmetallicity} from empirical relations
(see Methods; Table~\ref{tab:properties}), and detect low magnetic activity indicated by the presence of an H$_\alpha$ absorption line in the stellar spectrum.

%Compositions
The three exoplanets are among the smallest and nearest transiting exoplanets known to date (Fig.~\ref{fig:TOI-270_in_context}).
The radius of {\Ntarget}~b places it in a planetary population distinct from planets c and d; the trio is separated by the planetary radius gap around 1.7--2.0\,{\rearth} which divides two populations of planets, rocky super-Earths and gas-dominated sub-Neptunes (e.g. \cite{Fulton2018, VanEylen2018, Owen2013}; Fig.~\ref{fig:TOI-270_in_context}).
{\Nplanetb} likely falls into the regime of Earth-like/rocky compositions, while planets c and d are possibly water-ice or gas-dominated sub-Neptunes when employing statistically predicted masses of {\bMcompanionMearth}~{\mearth}, {\cMcompanionMearth}~{\mearth}, and {\dMcompanionMearth}~{\mearth}, respectively (\cite{Fortney2007,Owen2013,Chen2017}; Fig.~\ref{fig:TOI-270_in_context}). The diversity of the {\Ntarget} system thus provides an interesting case study for planet formation and photoevaporation, which can be driven by future observations using transit timining variations (TTVs), radial velocities (RVs) and transmission spectroscopy (discussed below).

\begin{figure*}[!htbp]
    \centering
    \begin{subfigure}[b]{0.65\textwidth} %0.71
        \centering
    \includegraphics[width=\columnwidth]{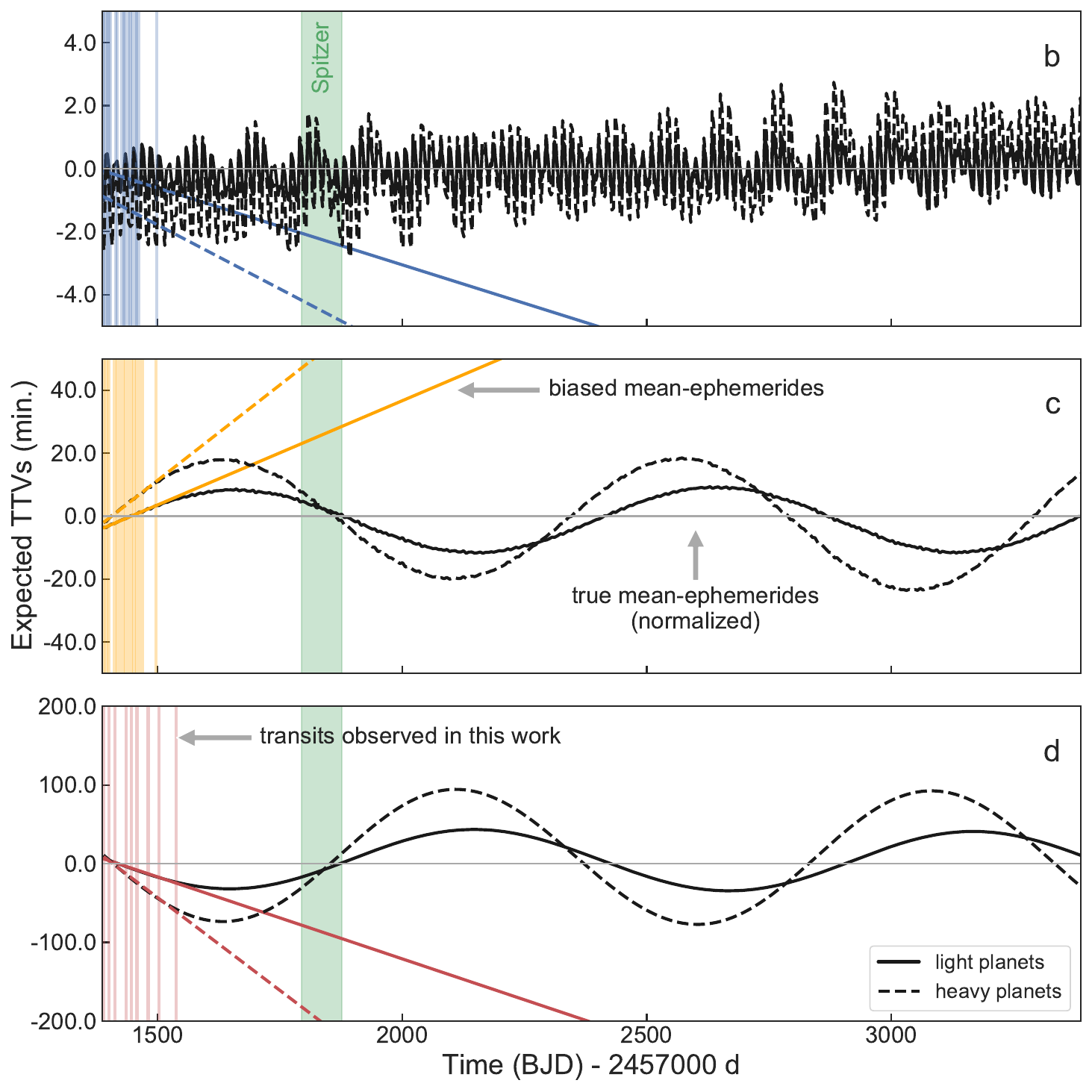}
    \end{subfigure}
    \begin{subfigure}[b]{0.34\textwidth} %0.28
        \centering
        \includegraphics[width=\columnwidth]{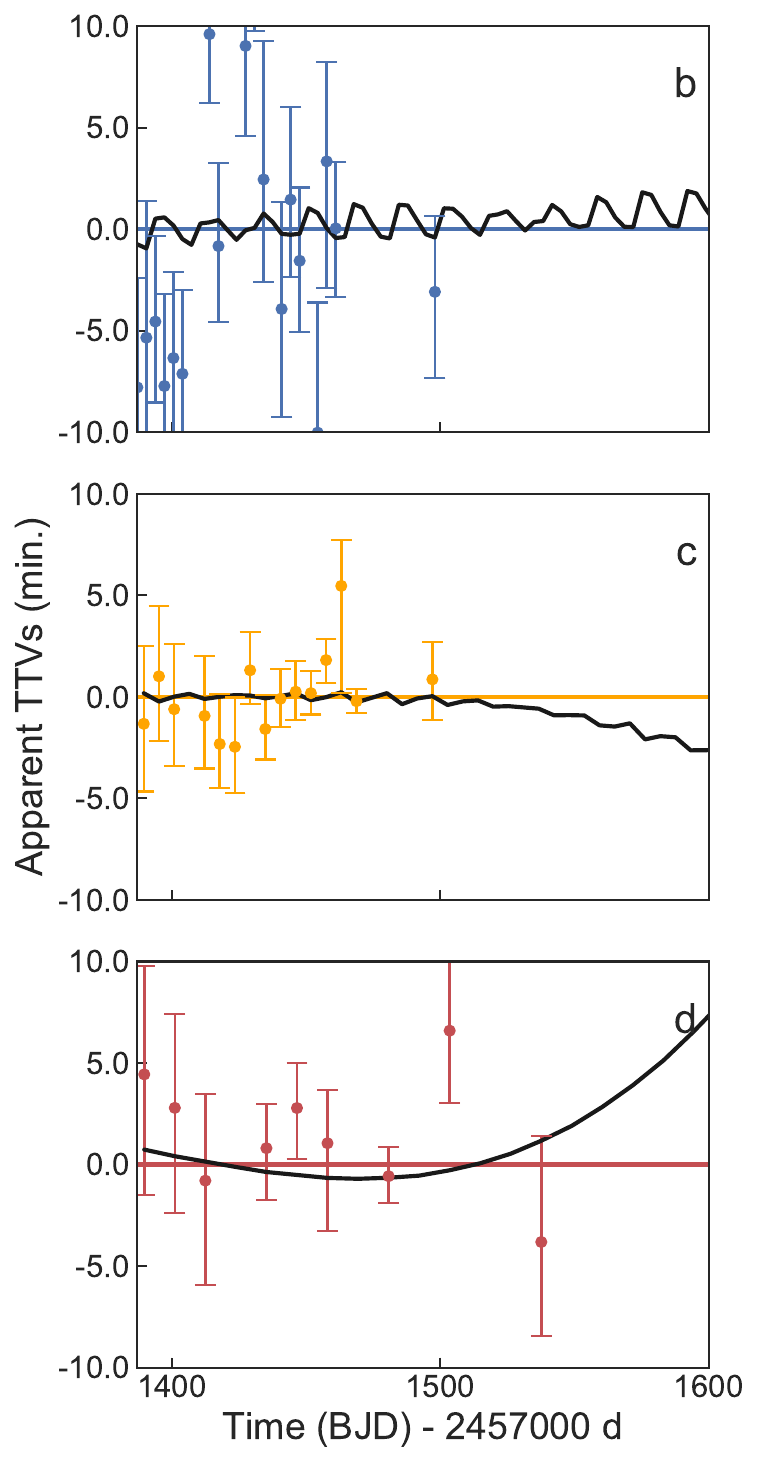}
    \end{subfigure}
    \caption{
        Expected and apparent transit timing variations (TTVs) of the {\Ntarget} system.
        \textit{Left}: Expected TTVs of the {\Ntarget} system, assuming the true mean-ephemerides of the system were known. Two example simulations show `light planets' (solid black lines) with masses predicted from \cite{Chen2017} and `heavy planets' (dashed black lines) composed of 50\% water ice and 50\% rock \cite{Zeng2016}. A future {\spitzer} observation window is marked in green.
        The expected TTVs have amplitudes of $\gtrsim10$\,min. (planet c) and $\gtrsim30$\,min. (planet d) and super-periods of 1000--1100\,days.
        However, the current observations (coloured vertical lines) span only a short, approximately linear part of the TTV signal, biasing mean-ephemerides fits (dashed and solid coloured lines).
        \textit{Right}: Apparent TTVs when the mean-ephemerides are predicted from only the observed transits (and thus biased). Coloured error bars show TTVs and 68.3\% credible intervals from the best-fit model, coloured lines show the best-fit mean-ephemerides, and solid black lines show again the simulation for `light planets'. 
        Hence, no transit timing variations are yet discernible in the combined {\tess} and follow-up data.
    }
    \label{fig:ttvs}
\end{figure*}

%TTVs and stability
Since the planets orbit near a resonant configuration, one can expect to measure TTVs - and thus planet masses - in the near future.
The proximity to the 2:1 resonance for {\Nplanetc} and d suggests that their perturbations lead to significant TTVs for both planets. 
If the inner planet pair (near 5:3 resonance) has a high relative eccentricity, it could also lead to observable TTVs for planet b.
However, given the available observation span of the {\tess} and follow-up data ($\sim$120~days), and the transit timing uncertainty ($\sim$2--5 minutes), {\Ntarget} is still best described as a multi-planet system on circular orbits with constant periods; we find no strong Bayesian evidence for eccentricity nor for TTVs (Supplementary Table~2).
We thus assess the theoretically expected amplitude and super-period of the TTVs through 4-body simulations \cite{Deck2014}. 
We find that the current observation span samples only a short and approximately linear part of the full TTV signal, which has a super-period of 1000-1100~days (Fig.~\ref{fig:ttvs}).
The TTV amplitudes of planets c and d are expected to be $\gtrsim10$\,min. and $\gtrsim30$\,min., respectively.
These are approximate lower limits, as the planet masses are currently predicted rather than measured, and the orbits are assumed to be circular.
Dynamical stability simulations show that the system is exceedingly stable for a range of eccentricities and planetary masses (see Supplementary Information), opening the possibility of non-circular orbits and even higher densities (and thus even larger TTVs).
Future transit observations sampling the super-period with moderate time-precision ($\sim$ few min.) should thus be sufficient to determine the TTV signal.
Importantly, the bias in predicting future transits from the linear ephemerides fit increases rapidly. For example, after just one year, planet d will have a systematic transit window offset by $>1$~hour due to the dynamical interactions. All this motivates the need for a continuous follow-up campaign, with observations every few months over the next 1--2 years.

%RV, TTV cont'd, and comparison to known multi-planet systems
Moreover, {\Ntarget} is inactive and much brighter than most comparable multi-planet hosts (especially in the infrared), making it a good target for precise radial velocity measurements with {\harps} or {\espresso}. 
We expect RV semi-amplitudes of around 2, 5, and 3\,m/s for planets b, c, and d, given the predicted masses.
This opens up the potential for accurate determination of the planets' masses and eccentricities in an independent and complementary way to TTV studies.
The majority of comparable multi-planet systems discovered by {\kepler} are too faint for RV follow-up (although K2 improved this situation to some degree).
Only few bright-enough systems comparable to {\Ntarget} are known, such as K2-3 \cite{Crossfield2015}, K2-18 \cite{Montet2015}, and LHS~1140 \cite{Dittmann2017}.
However, these and most {\kepler} systems are not as close to resonances, even when they do feature planets with similar sizes and orbital spacings \cite{Lissauer2011}.
Consequently, very few other multi-systems with small planets are as suitable as {\Ntarget} for complementary characterisation by both TTVs and RVs. 
This ultimately will provide insights into both the compositions and formations of three very interesting planets, which can be representative for compact multi-planet systems around M~dwarfs.

%JWST
As one of the nearest (and hence brightest) M~dwarf systems with transiting exoplanets, {\Ntarget} is also a promising target for atmospheric characterisation studies.
The low equilibrium temperatures of planets c and d ({\cTeq}~K and {\dTeq}~K) make them rare objects among currently known transiting super-Earth-sized and sub-Neptune-sized planets, and thus additionally compelling. 
Further, {\Ntarget} will be observable with the James Webb Space Telescope ({\jwst}) for 215 days per year, allowing easy observation scheduling.
Absorption features of planets c and d are expected to be readily detectable with $\mathrm{SNR} > 40$ and $\mathrm{SNR} > 60$, respectively, from just one transit with the NIRISS instrument and assuming cloud-free and H$_2$-dominated atmospheres \cite{Batalha2017,Kempton2018}.
Hence, {\Ntarget} provides a rare opportunity to test whether planets in compact multi-planet systems share the same formation history by comparing the atmospheric composition and thickness \cite{Hansen2013, Millholland2017}.
Moreover, it could be possible to constrain the ocean loss on these planets, by uniting the red-sensitive {\jwst} observations (probing $O_3$ abundances) with ground-based, visible-spectrum observations from the Extremely Large Telescopes (probing $O_2$ abundances; e.g. \cite{Luger2015, Serindag2019}).
Alternatively, one could search for $CO$ and $O_4$ features as indicators \cite{Schwieterman2016}.

%Habitability
The equilibrium temperature of {\Nplanetd} also places the planet within the survivable range of temperatures for extremophile organisms (Fig.~\ref{fig:TESS_lightcurve}, \cite{Takai2008}).
This temperate small exoplanet can thus be a unique laboratory, being exceptionally suited for characterisation by TTVs, RVs and transmission spectroscopy.
While planet d itself might not be rocky and potentially too massive for habitable oceans \cite{Noack2016}, plus it might suffer from runaway greenhouse effects and extreme water loss \cite{Luger2015}, it could host temperate rocky moons (see e.g. \cite{Kane2017}).
Additional companions beyond the orbit of {\Nplanetd} could also fall within the terrestrial-like habitable zone (0.10--0.28\,AU \cite{Kopparapu2014}) without impacting the stability of the system (see Supplementary Information).
Such planets are not expected to be detectable in the current data alone, encouraging future exploration (see Methods; Supplementary Fig.~7).
The host star, {\Ntarget}, is remarkably well suited for future habitability searches, as it is particularly quiet for an M~dwarf (e.g. \cite{Guenther2019}); it shows no signs of rotational variability, spots or stellar flares during our photometric observations, and low H-alpha activity in the reconnaissance spectra.
This makes it an ideal target for radial velocity surveys searching for additional planets in the habitable zone \cite{Newton2016}.
Moreover, the star is unlikely to sterilise or diminish the atmospheres of its planets through flaring or coronal mass ejections at its current stage (e.g. \cite{Lammer2007,Cohen2015,Tilley2019};
note that the star might have been more active at a younger age). 

We note two caveats: 
first, tidal effects can substantially influence localised habitability, which is not included in any terrestrial-like habitable zone definition.
In fact, due to the short orbital distances of the three planets, the planets rotation is expected to be tidally locked to their orbits.
Using dynamical simulations including planetary tides, we find that the obliquities decrease down to zero and all planetary rotation periods evolve towards pseudo-synchronisation in a timescale shorter than 10$^{5}$ years (see Supplementary Information). 
The second caveat is that the equilibrium temperature is not necessarily reflective of the surface temperature -- follow-up studies investigating the bulk masses, tidal locking, atmospheric compositions and pressures, and recirculation of gas and/or liquid water are required to determine any surface habitability.
Nevertheless, the {\Ntarget} system stands representative for a demographic of exoplanets in the potential habitable zone of M~dwarfs, paving the way for future habitable zone planet discoveries with {\tess}.

%Other TCEs
Compact multi-planet systems like {\Ntarget} are often accompanied by other small planets on short orbits. For example, this was the case for TRAPPIST-1, whose initial three planet signals were later found to be part of a seven planet resonance chain \cite{Gillon2017}.
We thus perform a search for additional components, and identify two more signals with a signal-to-noise ratio $\mathrm{SNR}>5$ (see Methods; Supplementary Table~4). 
However, after inspecting the data, we suggest that these are likely not planets but systematic artefacts. Nevertheless, long-period or non-transiting companions might accompany the planet trio, and could be detected with follow-up monitoring.

%Outlook
Soon, we will be able to precisely measure the masses of {\Nplanetc} and d through photometric follow-up of the significant TTVs caused by the multi-planet dynamics, and, independently, through RV observations enabled by the star's brightness and quietness. 
{\Ntarget} thus provides three new exoplanets which soon will fulfil the primary goal of the {\tess} mission (detecting and measuring the masses of at least 50 planets smaller than Neptune). 
Even more, we will be able to study the atmospheric composition of {\Ntarget}'s planets via transmission spectroscopy with {\jwst} and the ELTs with high SNR and near-optimal visibility.
Falling on both sides of the planet radius gap, the formation of these three interesting planets is likely representative of many other systems. 
All follow-up studies together (TTVs, RVs and transmission spectroscopy) will give insight into the bulk and atmospheric compositions of the planets, and provide an interesting case study for formation and photoevaporation.
Finally, with planet d falling into a temperate regime, and potentially more planets waiting to be discovered in the habitable zone, {\Ntarget} could provide an exemplary case for exoplanet habitability studies in the future.

% Figures

\clearpage
% >>   HZ_inner_edge = 0.10044 #in AU
% >>   HZ_outer_edge = 0.282 #in AU
% >>   Teq395_inner_edge = 0.05571900821346702 #in AU
% >>   Teq273_outer_edge = 0.11651857349044564 #in AU

\clearpage

\clearpage

\clearpage
% Tables
\newgeometry{left=0.5in, right=0.5in} 
\begin{table*}
\centering
\scriptsize
\caption{Properties of the {\Ntarget} system}
\begin{tabular*}{1\textwidth}{lcccc}
\hline
\hline
Parameter&\multicolumn{3}{c}{Value}&Source\\
\hline
\multirow{2}{*}{\textbf{Star}}&\multicolumn{3}{c}{\textbf{TOI~{\toiid}, TIC~{\ticid}, 2MASS~{\twomassid}}}&\\
 &\multicolumn{3}{c}{\textbf{Gaia~{\gaiaid}, L~231-32}}&\\
Right ascension, Declination (J2000)& \multicolumn{3}{c}{{\RA}, {\Dec}} & {\gaia} DR2	\\
Longitude, Latitude (ecl.; J2000) & \multicolumn{3}{c}{{\Longecl}, {\Latecl}} & via {\gaia} DR2	\\
\multirow{4}{*}{Magnitudes} & \multicolumn{3}{c}{ TESS-mag={\TESSmag} } & TICv7\\
 & \multicolumn{3}{c}{ V={\Vmag}, g={\gmag}, r={\rmag}, i={\imag} } & UCAC4\\
 & \multicolumn{3}{c}{ G={\GAIAmag}, $\mathrm{b_p}$={\GAIAbpmag}, $\mathrm{r_p}$={\GAIArpmag} } & {\gaia} DR2\\
 & \multicolumn{3}{c}{ J={\Jmag}, H={\Hmag}, K={\Kmag} } & 2MASS\\
Proper motion, $\mu_{{\rm R.A.}}$, $\mu_{{\rm Dec.}}$ ({\masy}) & \multicolumn{3}{c}{\propRA, \propDec } & {\gaia} DR2 \\
Parallax, $\varpi$ (mas)	& \multicolumn{3}{c}{{\parallax}} & {\gaia} DR2 \& \cite{Stassun2018b} \\   
Distance, $d_\star$ (parsec) & \multicolumn{3}{c}{{\distancepc}} & via {\gaia} DR2 \\  
Distance, $d_\star$ (ly) & \multicolumn{3}{c}{{\distancely}} & via {\gaia} DR2 \\
\multirow{2}{*}{Mass, {\mstar} ({\msun})} & \multicolumn{3}{c}{{\starmassBenedictsixteen}} &via ER$^a$\\ %\cite{Benedict2016} \\
   & \multicolumn{3}{c}{{\starmassMannnineteen}}           &via ER$^a$ \\ %\cite{Mann2019} \\
\multirow{2}{*}{Radius, {\rstar} ({\rsun})} & \multicolumn{3}{c}{{\starradiusBenedictsixteenBoyajiantwelve}}  & via ER$^a$ \\ % \cite{Benedict2016}\&\cite{Boyajian2012} \\
  & \multicolumn{3}{c}{{\starradiusMannfifteen}}          &via ER$^a$ \\ %\cite{Mann2015} \\
Density, $\rho_\star$ ({\gccc})   & \multicolumn{3}{c}{{\meanhostdensity}}                    &fit$^b$ \\
Luminosity, $L_\star$ ($L_\odot$)    & \multicolumn{3}{c}{{\starbolometricluminosity}}       &via ER$^a$ \\ %\cite{Mann2019} \\
Effective temperature, {\teff} (K)              & \multicolumn{3}{c}{{\starteff}}                       &via ER$^a$ \\ %\cite{Mann2019} \\
Metallicity, [Fe/H]              & \multicolumn{3}{c}{{\starmetallicity}}                       &via ER$^a$ \\ %\cite{Dittmann2016} \\
Spectral type            & \multicolumn{3}{c}{{\startype}}                       &via ER$^a$ \\ %\cite{Pecaut2013} \\
\hline
\textbf{Planets} & \textbf{TOI-270~b} & \textbf{TOI-270~c} & \textbf{TOI-270~d} &\\
Orbital period, $P$ &$3.360080_{-0.000070}^{+0.000065}$ & $5.660172\pm0.000035$ & $11.38014_{-0.00010}^{+0.00011}$ & fit\\
Mid-transit time, $T_0 - 2,457,000~\mathrm{(BJD_\mathrm{TDB})}$ & $1461.01464_{-0.00093}^{+0.00084}$ & $1463.08481\pm0.00025$ & $1469.33834_{-0.00046}^{+0.00052}$ & fit\\
Radius ratio, $R_p / R_\star$ & $0.0300_{-0.0011}^{+0.0015}$ & $0.05825_{-0.00058}^{+0.00079}$ & $0.05143\pm0.00074$ & fit\\
Sum of radii over semi-major axis, $(R_\star + R_p)/a$ & $0.0588_{-0.0046}^{+0.014}$ & $0.03919_{-0.00087}^{+0.0024}$ & $0.02530_{-0.00042}^{+0.00052}$ & fit\\
Cosine of orbital inclination, $\cos{i}$& $0.024_{-0.015}^{+0.024}$ & $0.0083_{-0.0051}^{+0.0073}$ & $0.0054_{-0.0027}^{+0.0021}$ & fit\\
Transit depth, $\delta$ (parts per thousand) & $0.901_{-0.066}^{+0.092}$ & $3.394_{-0.068}^{+0.094}$ & $2.645\pm0.078$ & derived\\
Stellar radius over semi-major axis, $R_\star/a$ & $0.0572_{-0.0045}^{+0.013}$ & $0.03703_{-0.00081}^{+0.0023}$ & $0.02406_{-0.00040}^{+0.00049}$ & derived\\
Planetary radius over semi-major axis, $R_p/a$ & $0.00170_{-0.00017}^{+0.00050}$ & $0.002154_{-0.000059}^{+0.00016}$ & $0.001237_{-0.000030}^{+0.000036}$ & derived\\
Planetary radius, $R_p$ (\rearth) & $1.247_{-0.083}^{+0.089}$ & $2.42\pm0.13$ & $2.13\pm0.12$ & derived\\
Orbital semi-major axis, a (\rsun) & $6.58_{-1.2}^{+0.71}$ & $10.14_{-0.71}^{+0.65}$ & $15.76\pm0.89$ & derived\\
Orbital semi-major axis, a (AU) & $0.0306_{-0.0057}^{+0.0033}$ & $0.0472_{-0.0033}^{+0.0030}$ & $0.0733\pm0.0042$ & derived\\
Orbital inclination, $i$ (degree) & $88.65_{-1.4}^{+0.85}$ & $89.53_{-0.42}^{+0.30}$ & $89.69_{-0.12}^{+0.16}$ & derived\\
Orbital eccentricity, $e$ & 0 & 0 & 0 & (fixed)\\
Impact parameter, $b$ & $0.41\pm0.26$ & $0.22_{-0.14}^{+0.17}$ & $0.224_{-0.11}^{+0.083}$ & derived \\ 
Total transit duration, $T_{1-4}$ (hours) & $1.387_{-0.034}^{+0.040}$ & $1.658_{-0.012}^{+0.015}$ & $2.148\pm0.018$ & derived \\
Full transit duration, $T_{2-3}$ (hours) & $1.278_{-0.038}^{+0.033}$ & $1.462_{-0.015}^{+0.011}$ & $1.927_{-0.022}^{+0.020}$ & derived \\
Equilibrium temperature, $T_\mathrm{eq}$ (K) & {\bTeq} & {\cTeq} & {\dTeq} & derived$^c$\\
\hline
\multicolumn{5}{l}{
\begin{minipage}{7.1in}
Listed values are the medians and 68.3\% credible intervals.
$^a$ derived using empirical relations (see Methods);
$^b$ fitted using a prior derived from the radius and mass (see Methods);
$^c$ derived using an albedo of 0.3 (Earth-like), and emissivity of 1.
\end{minipage}}\\
\end{tabular*}
\label{tab:properties}
\end{table*}
\restoregeometry

%%%%%%%%%%%%%% Main References %%%%%%%%%%%%%%%%
 \newcommand{\noop}[1]{}

%%%%%%%%%%%%%% Corresponding author, Acknowledgements, etc. %%%%%%%%%%%%%%

\clearpage

\subsection*{Correponding author}
Correspondence and requests for materials should be addressed to Maximilian N. G{\"u}nther (\url{maxgue@mit.edu}).

\subsection*{Acknowledgments}
%people
We thank Benjamin J. Fulton and Erik Petigura for providing their data to recreate the radius gap histogram in Fig.~\ref{fig:TOI-270_in_context}.
%TESS
Funding for the TESS mission is provided by NASA's Science Mission directorate.
We acknowledge the use of public TESS Alert data from pipelines at the TESS Science Office and at the TESS Science Processing Operations Center.
This research has made use of the Exoplanet Follow-up Observation Program website, which is operated by the California Institute of Technology, under contract with the National Aeronautics and Space Administration under the Exoplanet Exploration Program."
This paper includes data collected by the TESS mission, which are publicly available from the Mikulski Archive for Space Telescopes (MAST).
Resources supporting this work were provided by the NASA High-End Computing (HEC) Program through the NASA Advanced Supercomputing (NAS) Division at Ames Research Center for the production of the SPOC data products.
%Follow-up observations
TRAPPIST is funded by the Belgian Fund for Scientific Research (Fond National de la Recherche Scientifique, FNRS) under the grant FRFC 2.5.594.09.F, with the participation of the Swiss National Science Fundation (SNF). The research leading to these results has received funding from the ARC grant for Concerted Research Actions, financed by the Wallonia-Brussels Federation.
This work uses observations collected at the European Organisation for Astronomical Research in the Southern Hemisphere under ESO programme 102.C-0503(A).
%Gaia
This work makes use of results from the European Space Agency (ESA) space mission Gaia. Gaia data are being processed by the Gaia Data Processing and Analysis Consortium (DPAC). Funding for the DPAC is provided by national institutions, in particular the institutions participating in the Gaia MultiLateral Agreement (MLA).
%2MASS
This publication makes use of data products from the Two Micron All Sky Survey, which is a joint project of the University of Massachusetts and the Infrared Processing and Analysis Center/California Institute of Technology, funded by the National Aeronautics and Space Administration and the National Science Foundation.
%ADS
This research has made use of NASA's Astrophysics Data System Bibliographic Services.
%Exoplanetarchive
This research has made use of the NASA Exoplanet Archive, which is operated by the California Institute of Technology, under contract with the National Aeronautics and Space Administration under the Exoplanet Exploration Program.
%personal
MNG, CXH and JB acknowledge support from MIT's Kavli Institute as a Torres postdoctoral fellow. 
DD acknowledges support for this work provided by NASA through Hubble Fellowship grant HST-HF2-51372.001-A awarded by the Space Telescope Science Institute, which is operated by the Association of Universities for Research in Astronomy, Inc., for NASA, under contract NAS5-26555.
TD acknowledges support from MIT's Kavli Institute as a Kavli postdoctoral fellow.
Work by BTM was performed under contract with the Jet Propulsion Laboratory (JPL) funded by NASA through the Sagan Fellowship Program executed by the NASA Exoplanet Science Institute.
JGW is supported by a grant from the John Templeton Foundation. The opinions expressed in this publication are those of the authors and do not necessarily reflect the views of the John Templeton Foundation.
SW thanks the Heising-Simons Foundation for their generous support.
MG and EJ are FNRS Senior Research Associates.
KH acknowledges support from STFC grant ST/R000824/1.
B.R-A acknowledges funding support from CONICYT PAI/CONCURSO NACIONAL INSERCIÓN EN LA ACADEMIA, CONVOCATORIA 2015 79150050 and FONDECYT through grant 11181295.
JCS acknowledges funding support from Spanish public funds for research under projects ESP2017-87676-2-2 and RYC-2012-09913 ('Ram\'on y Cajal’ programme) of the Spanish Ministry of Science and Education. 

%Individual contributions
\subsection*{Individual contributions}
MNG: project lead, global analyses and interpretation;
FJP: TFOP SG1 (photometric follow-up), dynamic stability simulations, tides simulations;
JAD: stellar parameters, TTV simulations, ANU spectra analysis;
DD: global light curve analysis, TTV analysis, atmospheric characterisation prospects;
SRK: dynamic stability simulations, habitability prospects;
TD: global light curve analysis, TTV analysis;
ADF, BTM: TFOP SG2 (spectroscopy), FIRE spectra analysis, stellar parameters;
CH: TESS lightcurve analysis;
TDM: false positive analysis;
AB: stellar parameter analysis;
LGB, JB, JLB, MI, SNQ: TFOP SG2 (spectroscopic follow-up);
KAC, JDA, KB, KIC, TG, MG, KH, GI, EJ, JFK, FM, GM, EP, HMR, RS, JCS, TGT, AS, IAW: TFOP SG1 (photometric follow-up);
JJL: orbital dynamics, resonances, TTVs;
EM, DRC, IC, SBH, RAM, JS: TFOP SG3 (direct imaging follow-up);
AV: archival image analysis, TESS light curve analysis;
SW: global light curve analysis;
JGW: overluminous binary analysis, HR diagram analysis;
GRR, RKV, DWL, SS, JNW, JMJ: TESS architects;
NB, NG, SL, BRA: TESS TSO;
DAC, GF, EBT, JDT: TESS SPOC;
MF: TESS POC.

%% file: arxiv_methods.tex
\section*{Methods}

\subsection*{Discovery, follow-up and vetting}

{\Ntarget} was observed in {\tess} short (2\,min.) cadence mode in Sectors 3--5 (spanning 2018-Sep-20 to 2018-Dec-11; Supplementary Table~1).
The target was selected by the {\tess} Input Catalog \cite{Stassun2018} and Cool Dwarf Catalog \cite{Muirhead2018}.
The mission team alerted on three transiting exoplanet candidates in the {\tess} lightcurves in early December 2018 (Fig.~\ref{fig:TESS_lightcurve}). The lightcurves were extracted using the Science Processing Operations Center (SPOC) pipeline operated at the NASA Ames Research Center \cite{Jenkins2002, Jenkins2010, Smith2012, Stumpe2014, Jenkins2016, Jenkins2017}. Note that the TESS data coverage is interrupted for data downlink for circa 1 day every 13.7 days. Data from circa BJD$_\mathrm{TDB}$ 2,458,419 to 2,458,421 is masked during a period of thermal settling.
Thorough verification of the true nature of these transit events is crucial, because planet-like signals are frequently mimicked by systematic noise, or by astrophysical false positives (e.g \cite{Cameron2012}).
Super-Earth-sized and sub-Neptune-sized signals, in particular, are prone to be mimicked by background eclipsing binaries blended into the photometric aperture (e.g \cite{Guenther2017a}).
Moreover, constant light from unresolved background or companion stars can bias the interpretation by leading to an underestimation of planet radii (e.g \cite{Guenther2018}).
A discovery of three independent, periodic signals lends confidence, as multi-transit systems have a high probability of being real planets (see e.g. \cite{lissauer2012}).
In order to confidently rule out systematic noise and false positives, and to strengthen our hypothesis that the candidates are planets, we follow the subsequent candidate validation protocol. Our protocol is similar to the approaches for M~dwarf planet validation in the literature (e.g. \cite{Muirhead2014,Gillon2016,Vanderspek2019,Quinn2019}), and partly even more extensive than those.

All three candidates pass all the validation tests performed by SPOC Data Validation module \cite{Twicken2018, Li2019}. These include an odd/even transit fit test against eclipsing binaries, a search for weak secondaries at the same period, a ghost diagnostic test against background eclipsing binaries and scattered light, and the difference image centroid test (a powerful and sensitive test of whether the source of the transit-like features are coincident with the target star). For all three planets, the transit source is displaced from the out-of-transit centroid by no more than 2 arcsec at the 0.6 sigma level, well within the 3 sigma confusion radius. Candidate 1 fails the weak secondary test at the 8.1 sigma level at a phase offset of 0.04 from the primary transit, but this feature is actually a transit of the second candidate and hence, can be ignored. In addition, the statistical bootstrap test quantifies the probability that the signal is a false alarm due to statistical fluctuations in the light curve, which is $<5\times 10^{-26}$ in all three cases.

Independently, we test for background objects that could influence our observations to-date. We study the {\tess} centroid time series and the image pixels during transits, search for known {\gaia} DR2 sources in the photometric aperture, and inspect archival images and photographic plates from 1983 until present (since {\Ntarget} has a high proper motion; Fig~\ref{fig:archival_images}). Neither of these analyses indicate signs of background objects.
Additionally, the {\gaia} DR2 photometric excess noise is 0, which was suggested to rule out faint blends down to 1'', and bright blends down to 0.1'' separation \cite{Rizzuto2018}.

We also test whether {\Ntarget} itself could be an unresolved equal-magnitude binary. We find that the photometric \cite{Winters2015} and trigonometric distances \cite{Gaia2018} of {\Ntarget} agree, which would have differed by a factor of $\sqrt{2}$ for binaries with similar brightness. The target also lies separated from known multi-star systems on an observational Hertzsprung-Russel diagram, created using data from \cite{Winters2019}. Both findings effectively exclude this false positive scenario.

Next, we coordinated ground-based follow-up observations through the {\tess} Follow-up Observing Program (TFOP) working groups (\url{https://tess.mit.edu/followup/}; Supplementary Table~1; Supplementary Fig.~2).
The TFOP \textit{Seeing-Limited Photometry} sub group (SG-1) observed the target star with various ground-based facilities at the predicted transit times, searching for deep eclipses in nearby stars within a radius of 2.5', and finding all transits to be on {\Ntarget}.
Observations in different filters show no chromatic behaviour (i.e. no variation of transit depth), which would have indicated eclipsing binaries or blended stellar companions.
The involved facilities are: Las Cumbres Observatory (LCO) telescope network \cite{brown2013}; TRAPPIST-South (TS) \cite{jehin2011}; Siding Spring Observatory T17 (SSO T17); The Perth Exoplanet Survey Telescope (PEST); Mt. Kent Observatory (MKO-CDK700); and Myers-Siding Spring (Myers).

Moreover, the TFOP \textit{Recon Spectroscopy} sub group (SG-2) obtained two low- to medium-resolution spectra. Both are consistent with a single, isolated star, showing no signs of a composite spectrum (i.e. no double-lined binary).
The first spectrum was taken with the Folded-port InfraRed Echellete (FIRE) spectrograph on the 6.5\, Baade Magellan telescope at Las Campanas observatory, covering the 0.8--2.5 micron band with a spectral resolution of $R = 6000$. Using the empirical relations of \cite{newton15}, we estimate the following stellar parameters from the spectrum: $T_{eff} = 3706 \pm 76$\,K, $R = 0.44 \pm 0.028$\,{\rsun} and $L = 0.033 \pm 0.005\,L_\odot$. These values are consistent within $\sim$ 2 sigma with those derived through empirical photometric relations (Table~\ref{tab:properties}, and see below).
The second spectrum was taken with the Echelle Spectrograph on the Australia National University (ANU) 2.3\,m telescope, covering the wavelength region of $3900-6700\,\text{\AA}$ with a spectral resolution of $R = 23000$.

Finally, the TFOP \textit{High-resolution Imaging} sub group (SG-3) collected high-resolution adaptive optics images of the target with VLT/NaCo \cite{Lenzen2003, Rousset2003}. We collected nine exposures of 20s each with the $Ks$ filter and processed the data following a standard procedure, to obtain a clean image of the target (Supplementary Fig.~3).
To calculate the sensitivity to companions, we inject copies of the central source at varying angles and separations, and scale their brightness such that they could be detected at 5 sigma sensitivity with standard aperture photometry. No visual companions appear anywhere within the field of view, and the star appears single to the limit of our resolution (full-width at half maximum of $\sim$90\,mas).

In summary, based on physics and observations alone, we can reject all sources of systematic false alarms or astrophysical false positives but one; the only remaining physically possible scenario would be that {\Ntarget} itself were a hierarchical multi-star system. While it is highly implausible that faint stellar companions would mimic a planet signal with a period that matches the near-resonance of a multi-planet system, we still investigate this scenario. We independently validate the planet with \texttt{vespa} \cite{Morton2012, Morton2015code}, which validated over a thousand {\kepler} planets, and find a false positive probability of $<10^{-6}$. Note that this is even an overestimated upper limit, as this calculation only considers the {\tess} lightcurve and does not take into account the image-level data, multi-planet nature, and most follow-up information.

Altogether, these results validate the real planetary nature of the {\Ntarget} system.

\subsection*{Stellar parameters}

We retrieve stellar parameters such as coordinates, parallax and photometric magnitudes from the {\gaia} DR2, 2MASS, and UCAC4 catalogues (Table~\ref{tab:properties}). 
We correct the {\gaia} DR2 parallax for the systematic offset reported by \cite{StassunTorres2018}. Note, however, that e.g. \cite{Jao2016} suggest that objects closer than 25 pc have an even larger offset (based on {\gaia} DR1 data).
The stellar mass {\mstar}, radius {\rstar} and effective temperature {\teff} are estimated using empirical relations.
First, we translate the apparent K-band magnitude $m_\mathrm{K}$ into the absolute K-band magnitude $M_\mathrm{K}$ using the parallax, leading to $M_\mathrm{K}=$~{\Kmagabs}.
Next, we use the empirical relations given by Eq.~11 and Table~13 from \cite{Benedict2016} to calculate the mass, resulting in $M_\star=$~{\starmassBenedictsixteen}~{\msun}. 
Here, we assume a conservative error of 5\% to account for the scatter in the data from which the empirical relation is derived.
We then follow the empirical mass-radius relationship of Eq.~10 from \cite{Boyajian2012} and compute the stellar radius \rstar from this mass.
This gives $R_\star=$~{\starradiusBenedictsixteenBoyajiantwelve}~{\rsun}, again with a conservative estimate of a 5\% error.

For comparison, we additionally calculate the mass using the empirical relation provided by Eq.~2 and Table~6 from \cite{Mann2019}.
This results in $M_\star=$~{\starmassMannnineteen}~{\msun} (5\% error estimate). 
This mass is 10\% smaller than the one calculated above, agreeing only within 1.4~sigma of the estimated uncertainty.
Moreover, we compare the radius from above with the one gained via the empirical relations given in Table~1 from \cite{Mann2015},
leading to $R_\star=$~{\starradiusMannfifteen}~{\rsun} (5\% error estimate). 
This radius is 5\% smaller than the value calculated above, thus consistent within 0.7~sigma of the estimated uncertainty.

Overall, this underlines the necessity of more conservative error estimates, such as the ones we follow here. For completeness, all values are reported in Table~\ref{tab:properties}. From this point onward, we use the values computed from the relations of \cite{Benedict2016} and \cite{Boyajian2012} with estimated 5\% errors.

To estimate the stellar effective temperature {\teff} and spectral type, we first compute the bolometric correction in the K-band, $BC_\mathrm{K}$, following Table~3 from \cite{Mann2015}.
For this, the colour $V-J$ is calculated from the given magnitudes. 
We find $BC_\mathrm{K}=$~{\starBCK} (assuming 5\% errors).
This leads to a bolometric magnitude of 
$M_\mathrm{bol}=$~{\starbolometricmagnitude}.
From this, we calculate the bolometric luminosity 
$L=$~{\starbolometricluminosity}~$L_\odot$.
Finally, we compute {\teff} using the Stefan-Boltzmann law to be \teff~=~\starteff~K.
This corresponds to an {\startype} spectral type \cite{Pecaut2013} (\url{http://www.pas.rochester.edu/~emamajek/EEM_dwarf_UBVIJHK_colors_Teff.txt}, online 2019 Jan 20).

We estimate the metallicity of {\Ntarget} using the photometric method from \cite{Dittmann2016}. Due to molecular line blanketing in the optical regions of the spectrum, a more metal rich M~dwarf at a given absolute $K_s$ magnitude (i.e. mass) will have a redder colour than a more metal poor star of the same mass. While \cite{Dittmann2016} utilised $MEarth$-$K_s$ as the colour for their calibration, the {\it MEarth} bandpass is broadly similar to the $i$-band; hence, a similar relation can be calibrated with $i$ - $K_s$ (Dittmann et al., in prep.). Using this calibration, we estimate the metallicity of {\Ntarget} to be [Fe/H] = {\starmetallicity}. 

With the current observations, we cannot constrain the age of the star with certainty. Nevertheless, the absence of photometric variability, the low activity (shallow H$\alpha$ absorption feature), and the lack of Ca H/K emission lines in our spectra suggest the star is an old main sequence M~dwarf star \cite{Mamajek2008, Newton2016b, Newton2018}.

\subsection*{Modelling the data with allesfitter}

\texttt{allesfitter} \cite{allesfitter} (\url{https://github.com/MNGuenther/allesfitter}) is a publicly available, user-friendly software to model photometric and RV data. Its generative model encompasses multiple exoplanets, multi-star systems, star spots, and stellar flares.
For this, it provides one framework uniting the versatile packages 
\texttt{ellc} (light curve and RV models; \cite{Maxted2016}), 
\texttt{aflare} (flare model; \cite{Davenport2014}),
\texttt{dynesty} (static and dynamic nested sampling; \cite{Speagle2019}),
\texttt{emcee} (Markov Chain Monte Carlo sampling; \cite{Foreman-Mackey2013}) and 
\texttt{celerite} (GP models; \cite{Foreman-Mackey2017}).

Facing three Sectors of {\tess} data and $16$ follow-up lightcurves, the number of free parameters adds up to $\geq95$ (without considering TTVs).
This accounts for the following:
for each of the three planetary systems, there are at least five parameters (assuming circular orbits; seven parameters for eccentric orbits);
for each of the 16 instruments, there are two limb darkening parameters (for quadratic laws), 
one error scaling parameter (for white noise scaling), 
and two hyperparameters for the Gaussian Process kernel ($\sigma$ and $\rho$, see above).
Additionally including a free TTV offset parameter for every transit would lead to $\geq105$ free parameters (if only TTVs of planet d are considered; $\geq138$ for {\Ntransits} transits in total).
Neither Markov Chain Monte Carlo (MCMC) nor Nested Sampling are suited to reliably account for all the covariances in such high dimensionality (the `curse of dimensionality').

A common approach to bypass this high dimensionality is to fix certain nuisance parameters (e.g. limb darkening, errors and baselines) to pre-determined values and fit a strictly periodic global model. 
Next, to search for TTVs, each individual transit is fitted again while only the epoch is free and all other parameters are fixed.
The difference of the individual epoch fits from the global fit is recorded and interpreted as TTVs.
However, this approach neglects the covariances between physical parameters; for example, strong TTVs could bias the period, inclination and planet radius in a global fit. Fixing these parameters could result in forcing a `wrong template' onto the transit, thus biasing the extracted TTVs.

We try to improve this method and opt for the following seven-step approach:
\begin{enumerate}
\item To refine the transit locations reported by the SPOC pipeline, we perform a preliminary fit of the TESS Sectors 3--5 lightcurves using wide uniform priors\footnote{Note that all priors used in this work are additionally truncated to physical lower and upper bounds. None of the priors are unbounded, and the likelihood functions for all models converge to 0 as the model deviates from the data. All priors are jointly proper, ensuring posterior propriety.}.
\item We mask out an 8h window around every transit midpoint and fit for the noise and GP hyperparameters in the out-of-transit data of the TESS Sectors 3--5 lightcurves.
\item We propagate the out-of-transit posteriors of the noise and GP hyperparameters (from step 2) as priors into a fit of the in-transit-data of TESS Sectors 3--5. All planet and orbit parameters are sampled from wide uniform priors. 
\item Next, we turn our attention to the follow-up data sets. To gain information about the noise length scale correlated to the airmass trend, we fit the measured airmass curve of each observation with a GP model.
\item Many follow-up observations served the purpose of ruling out false positives such as blended eclipsing binaries; not all follow-up facilities have the ability to refine the transit parameters measured by TESS. As a consequence, several of these observations are dominated by red noise, on scales larger than the transit signals. To select only useful observations, we fit each follow-up data set separately, and each with two models: a `transit and noise' model, and a `noise-dominated' model. We record the Bayesian evidence $Z_\mathrm{transit}$ for each fit.
\begin{enumerate}
    \item `transit and noise' model: We propagate the posteriors of the planet and orbit parameters (from step 3) and of the GP time scale hyperparameter ($\rho$; from step 3) as priors into this fit. The remaining nuisance parameters (limb darkening, noise, and GP baseline) are sampled from wide uniform priors. 
    \item `noise-dominated' model: We fit a pure noise model. The posterior of the GP time scale hyperparameter ($\rho$; from step 3) is propagated as a prior into the fit. The noise and GP amplitude are sampled from wide uniform priors.
\end{enumerate}
\item For each follow-up observation, we compute the Bayes factor as $\Delta \ln{Z} = \ln{Z_\mathrm{transit}} - \ln{Z_\mathrm{noise}}$. There is strong evidence for a transit signal being recovered if $\Delta \ln{Z}>3$ \cite{Kass1995}. For the global analysis, we only include observations that fulfil this criterion (see Supplementary Table~1;
Supplementary Fig.~2).
We note that care must be taken with this criterion, as also white-noise dominated observations could get rejected, leading to biases in the transit depth. The theoretically ideal approach would be to fit all data simultaneously with appropriate red noise models.
\item Finally, we perform a global model of all data sets. For this, all nuisance parameters (limb darkening, noise, and GP baseline) are fixed to the median posterior of their individual fits. Wide uniform priors are set for the physical parameters. We perform this step in nine separate ways:
\begin{enumerate}
\item without TTVS, circular orbits
\item with TTVs for planet b, circular orbits
\item with TTVs for planet c, circular orbits
\item with TTVs for planet d, circular orbits
\item with TTVs for all planets, circular orbits
\item free eccentricity for planet b, no TTVs
\item free eccentricity for planet c, no TTVs
\item free eccentricity for planet d, no TTVs
\item free eccentricity for all planets, no TTVs
\end{enumerate}
\end{enumerate}

This approach makes use of the Bayesian laws by propagating information via priors wherever possible, but still has to neglect certain covariances between nuisance parameters. Under the assumption that different observations are independent and identically distributed, we argue that the impact of this on the global posteriors of the planet and orbit parameters (which are evaluated globally) is negligible. For example, we find no significant covariance between the limb darkening and the transit parameters in all our independent fits of individual data sets.

Moreover, we pass the stellar radius and mass reported in Table~\ref{tab:properties} as input into \texttt{allesfitter} to compute an (approximate) normal prior on the stellar density of $\rho_\star^\prime = 10.3 \pm 1.9$. Independent of that, the stellar density is calculated at each Nested Sampling step from the fitted parameters via $\rho_\star \approx \frac{3 \pi}{G P^{2}}\left(\frac{a}{R_{\star}}\right)^{3}$ \cite{Seager2003}. Then, the prior stellar density and the calculated stellar density are compared, and the fit gets penalised for discrepancies. The resulting best fit posterior gives $\rho_\star = 10.5 ^{+1.4}_{-2.2}$, agreeing with the prior.

The effects of potential planet atmospheres on the transit depth per band pass are negligible, and we thus fit for a single parameter (radius ratio) per planet for all photometric bands. Assuming the most extreme case of a low-density sub-Neptune with a rocky core and H$_2$-dominated atmosphere, the maximum transit depth change is $\Delta \delta < 100~\mathrm{ppm}$. In comparison, the 1 sigma statistical uncertainty on our measured transit depths for the three planets are $\sigma_\delta \sim 70-100~\mathrm{ppm}$.

To ensure we are not missing a TTV detection, we also independently model the system using various other codes and approaches, including the ones implemented in the \texttt{ExoFastv2} \cite{Eastman2017} package. All results are consistent with our \texttt{allesfitter} analysis, suggesting that the TTVs can currently not be detected given the $\sim$2--5 minute uncertainties on the transit timing in the data (Supplementary Table~2). 
The resulting lightcurves from the favoured model (no TTVs and circular orbits) are shown in Fig.~\ref{fig:TESS_lightcurve}, its physical parameters in Table~\ref{tab:properties}, and its posterior distributions in Supplementary Fig.~4.
We also report all nuisance parameters in Supplementary Table~3.

\subsection*{Searching for additional exoplanet candidates}

We search for additional threshold crossing events that might have not been detected by the automated pipeline.
Using the short-cadence data from Sectors~3--5, we first detrend the Pre-Search Data Conditioning Simple Aperture (PDC-SAP) flux using a Gaussian Process with a Matern 3/2-kernel to remove any remaining long-term systematics that could impact the transit search.
For the transit search, we use the software \texttt{transit least squares} (\texttt{TLS}) \cite{Hippke2019}.
The code is similar to the widely used transit search algorithm \texttt{box least squares} \cite{Kovacs2002}.
However, instead of fitting a box model to mimic transit dips in the lightcurve, \texttt{TLS} uses a physical transit model \cite{Mandel2002,Kreidberg2015}, to increase its detection efficiency.
The software also allows to include the stellar mass and radius from Table~\ref{tab:properties} as priors to generate the transit models.
We search for signals with periods between $\sim$0.2 and $\sim$40~days that exceed a signal-to-noise ratio $\mathrm{SNR}\geq5$.
This approach detects the initial threshold crossing events of the {\tess} pipeline with $\mathrm{SNR}>7$, and finds three additional threshold crossing events with an SNR between 5 and 7 (Supplementary Table~4). %\ref{tab:TLS}).
After careful vetting, we conclude that these are likely not planets, but systematic artefacts.

We also investigate the system's potential for hosting transiting planets in the terrestrial-like habitable zone, which could not have been discovered with the current observations alone. The habitable zone limits ranging 0.1--0.28\,AU correspond to periods between 18 and 85 days, and the TESS baseline covers circa 80 days. To quantify which transiting exoplanets might remain to be discovered (for example because of low SNR or data gaps), we injected planet signals into the TESS light curves with varying planet sizes and periods, ranging 1--2\,{\rearth} and 15--85 days. We then tried to recover these signals with a \texttt{TLS} search. We considered a signal to be recovered if a detected epoch matched the injected epoch to within one hour, and if a detected period matched any multiple of half the injected period to better than 5\%. We find that the regime of small exoplanets at orbits longer than $\sim$30\,days remains widely open for future exploration (Supplementary Fig.~7). %\ref{fig:injected_transit_search}). 
Hence, transiting long-period or non-transiting planets might still be waiting to be discovered in the terrestrial-like habitable zone.

\clearpage

%Facilities
\noindent\textbf{Facilities}\\
{\tess} \cite{Ricker2014}; 
Las Cumbres Observatory (LCO) telescope network \cite{brown2013};
TRAPPIST-South (TS) \cite{jehin2011};
Siding Spring Observatory T17 (SSO T17);
The Perth Exoplanet Survey Telescope (PEST); 
Mt. Kent Observatory (MKO-CDK700);
Myers-Siding Spring (Myers);
Magellan Folded-port InfraRed Echellette (FIRE) \cite{Simcoe2013};
Australia National University (ANU) Echelle spectrograph;
VLT NAOS-CONICA (NaCo) \cite{Lenzen2003, Rousset2003};

%Archival data \& catalogs
\noindent\textbf{Catalogs \& all sky survey archives}\\
TICv7 \cite{Stassun2018};
{\twomass} \cite{2MASS};
{\gaia} \cite{GAIA, Gaia2018}; 
UCAC4 \cite{UCAC4}.

%Data availability
\noindent\textbf{Data availability}\\
The {\tess} data analysed during the current study are available in the Mikulski Archive for Space Telescopes (MAST; \url{http://archive.stsci.edu}). 
The VLT NAOS-CONICA data analysed during the current study are available in the European Southern Observatory archive (\url{http://archive.eso.org}).
All other data sets analysed and/or generated during the current study are available from MNG on reasonable request.

%Code availability
\noindent\textbf{Code availability}\\
The following code used during the current study is publicly available (access information given in the respective references):
\texttt{python} \cite{Rossum1995},
\texttt{numpy} \cite{vanderWalt2011},
\texttt{scipy} \cite{Jones2001},
\texttt{matplotlib} \cite{Hunter2007},
\texttt{tqdm} (doi:10.5281/zenodo.1468033),
\texttt{seaborn} (\url{https://seaborn.pydata.org/index.html}),
\texttt{allesfitter} \cite{allesfitter},
\texttt{ellc} \cite{Maxted2016},
\texttt{aflare} \cite{Davenport2014},
\texttt{dynesty} \cite{Speagle2019},
\texttt{corner} \cite{Foreman-Mackey2016},
\texttt{TESS Transit Finder}, a customised version of the \texttt{Tapir} software package \cite{jensen2013},
{\tt AstroImageJ} \cite{collins2017},
\texttt{ttvfast} \cite{Deck2014},
\texttt{REBOUND} \cite{rein2012}.
Any customised scripts build on these codes are available from MNG on reasonable request.

%Competing interests
\noindent\textbf{Competing interests}\\
The authors declare no competing interests.

%% file: arxiv_supplement.tex
\renewcommand{\figurename}{Supplementary Figure}
\renewcommand{\tablename}{Supplementary Table}
\setcounter{figure}{0}
\setcounter{table}{0}

\section*{Supplementary Information}

\subsection*{On Bayesian statistics, Nested Sampling and Gaussian Processes}

Here, we briefly outline the key concepts of Bayesian statistics, Nested Sampling and Gaussian Processes, which we extensively use for all analyses.
Following Bayes' theorem, the `posterior' $P(\theta|M, D)$ is the degree of belief about the model $M$ and its parameters $\theta$ , which is updated based on data $D$. It is given by:
\begin{equation}
P(\theta|M, D) = \frac{P(D|\theta, M) P(\theta| M)}{P(D|M)}.
\end{equation}
Therein, the `likelihood' $P(D|\theta, M)$ is the probability of observing the data given the model and parameters. The `prior' $P(\theta| M)$ limits and informs the model parameters.
The last term, $P(D|M)$, is the `Bayesian evidence',
\begin{equation}
P(D|M) = \int P(D|\theta, M) P(\theta| M) \mathrm{d} \theta.
\end{equation}
and quantifies the degree of belief about the model itself given the data (marginalised over all parameters).
Comparing different physical models, which is often desired in exoplanet studies, relies on the estimation of the Bayesian evidence, $P(D|M)$.

Nested Sampling \cite{Skilling2004} is designed to directly compute the Bayesian evidence -- making it distinct from Markov Chain Monte Carlo (MCMC) approaches, which bypass this step.
For example, this enables the robust comparison of models with different numbers of exoplanets \cite{Hall2018}, circular versus eccentric orbits, or TTVs versus no TTVs.
With Nested Sampling we draw samples from the prior volume (of the model parameter space) with hard likelihood thresholds. Successively, samples with the smallest likelihood get rejected, until the posterior distribution is found.

For modelling correlated noise in the data, we employ a Gaussian Process (GP) jointly with our transit model fit.
A GP uses different kernels and metrics to evaluate the correlation between data points.
The squared distance $r^2$ between data points $x_i$ and $x_j$ is evaluated for any metric M as
\begin{align}
r^2 = ( x_i - x_j )^T M^{-1} ( x_i - x_j ).
\end{align}
We choose our GP with a series approximation of a `Matern 3/2 kernel' $k(r)$ using the \texttt{celerite} implementation \cite{Foreman-Mackey2017}:
\begin{align}
\begin{split}
    \small
    k (r) = \sigma^2 &\left[ (1+1/\epsilon) e^{-(1-\epsilon)\sqrt{3}r/\rho} \right. \\ 
                        &\left. \cdot (1-1/\epsilon) e^{-(1+\epsilon)\sqrt{3}r/\rho} \right].
\end{split}
\end{align}
This kernel has two hyperparameters that are fitted for: the amplitude $\sigma$, and the time scale $\rho$ of the correlations.
In this expression used by \texttt{celerite}, $\epsilon$ controls the quality of the series approximation and is set to $0.01$; in the limit $\epsilon \rightarrow 0$ it becomes the Matern-3/2 function.
This kernel can describe variations with a smooth, characteristic length scale together with rougher (i.e. more stochastic) features.

\subsection*{Orbital dynamics}

\begin{figure*}[!htbp]
    \centering
    \includegraphics[width=0.5\textwidth]{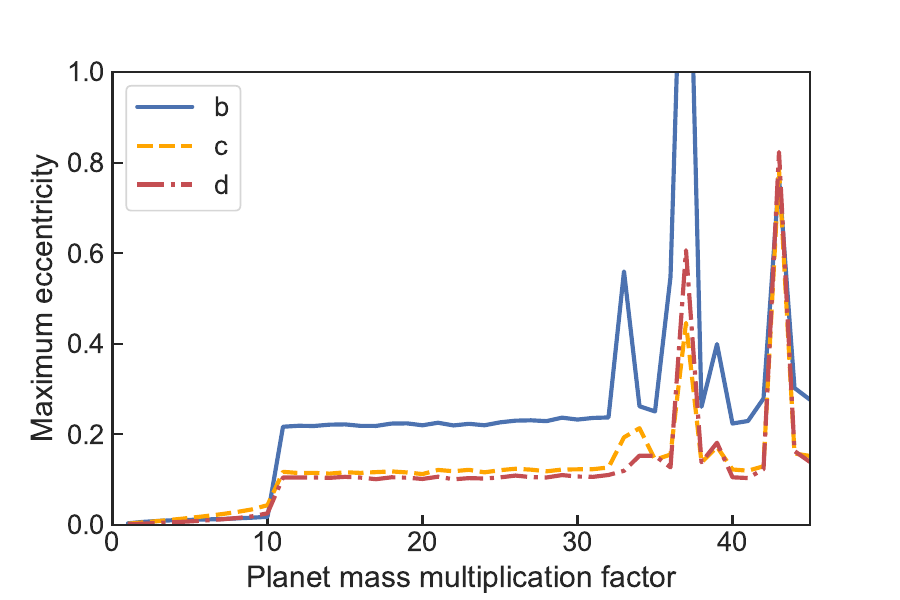}
    \caption{Dynamical analysis based on the Mercury Integrator, showing the planets' eccentricities over a range of masses (the predicted mass multiplied by a factor). The system is stable with eccentricities remaining below 0.05 for masses up to ten times the predicted mass. For masses 10--30 times higher, the system achieves stability but the interaction between planets begins to drive high eccentricities. At $\sim$30 times the original masses, the system would be chaotic.}
    \label{fig:stability_mass}
\end{figure*}

To investigate the dynamical stability of the {\Ntarget} system for a range of planet masses, we utilised the Mercury Integrator Package written by \cite{chambers1999}. The 4-body integrations were carried out for a duration of $10^6$ simulation years, equivalent to $1.1 \times 10^8$ orbits of the inner planet and $3.2 \times 10^7$ orbits of the outer planet. To ensure a sufficient time resolution, we adopt the criteria of \cite{duncan1998} and choose a time resolution of 0.05~days. Regarding the initial orbital conditions of the planets, we assume zero eccentricity, a periastron argument of $\omega = 90\degree$, and specify the time of inferior conjunction using the $T_0$ values for each of the planets shown in Table~1.%\ref{tab:properties}. 
The planet masses are adopted from the predicted values. We conduct a series of dynamical simulations that vary the mean anomaly (starting locations) for each of the planets. This technique explores the orbital parameter space that determines dynamical stability as a function of various system parameters \cite{kane2015,kane2016}.
Assuming initial circular orbits, we find that the system is exceedingly stable with eccentricities remaining below 0.4\% (Supplementary Fig.~\ref{fig:stability_mass}).
Gradually raising the assumed masses, we find that the system remains stable up to ten times the original mass estimates. In the range of 10--30 times the original masses, the system achieves stability but the interaction between planets begins to drive relatively high eccentricities and rapid angular momentum transfer between the orbits. Beyond $\sim$30 times the original masses, instability in the system becomes inevitable with planets either being ejected from the system or colliding with the host star.

\begin{figure*}[!htbp]
    \centering
    \includegraphics[width=1\textwidth]{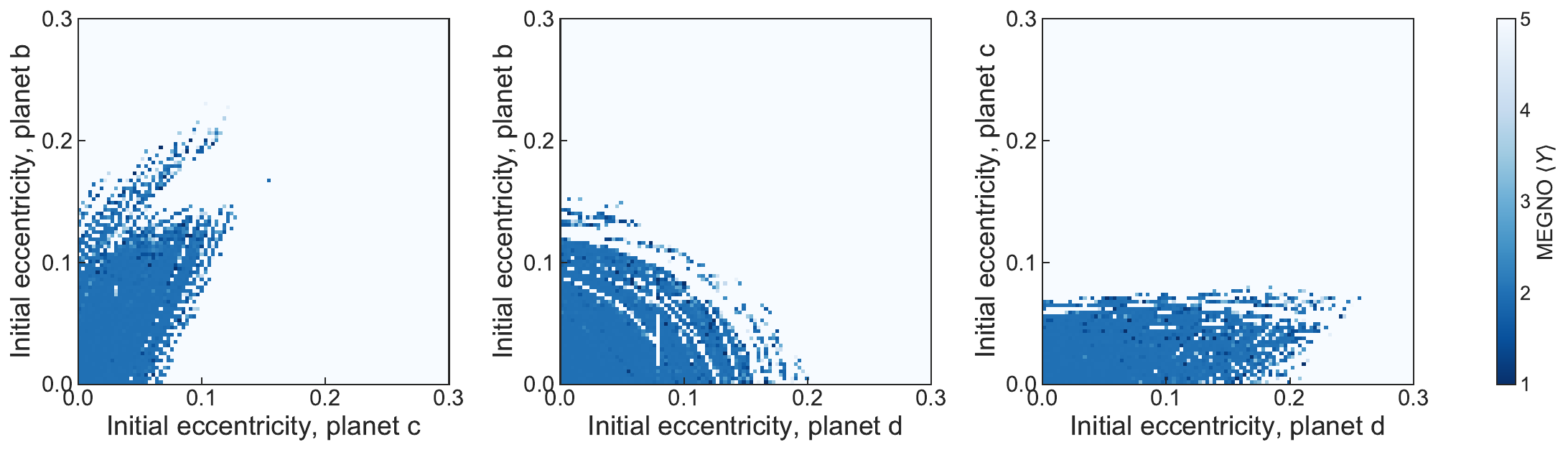}
    \caption{Dynamical analysis based on MEGNO-maps. The configurations are as follow: Left, free eccentricities e$_{b}$ and e$_{c}$ in the range of 0 to 0.3, while e$_{d}$=0. Middle, free e$_{b}$ and e$_{d}$, while e$_{c}$=0. Right, free e$_{c}$ and e$_{d}$, while e$_{b}$=0. All other planetary parameters are fixed. In all cases: 
    $\langle Y(t) \rangle \rightarrow 2$ for quasi-periodic orbits and $\langle Y(t) \rangle \rightarrow 5$ for chaotic systems. 
    This shows that the system is stable for a range of low eccentricities.}
    \label{fig:megno-maps}
\end{figure*}

Independently, to explore the system's stability in the context of non-circular orbits we computed the Mean Exponential Growth factor of Nearby Orbits, $Y(t)$ (MEGNO, \cite{cincottasimo1999,cincottasimo2000,cincotta2003}). This chaos index evaluates the stability of the bodies' trajectories after small perturbations. Each body's six-dimensional displacement vector, $\delta_{i}$, (position and velocity) is a dynamical variable from its `shadow particle' (a particle with slightly perturbed initial conditions). We obtained differential equations for each $\delta_{i}$ by applying a variational principle to the trajectories of the original bodies. Next, the MEGNO was computed from the variations as: 
\begin{equation}
 Y(t)=\frac{2}{t}\int_{0}^{t} \frac{\Arrowvert \dot{\delta}(s) \Arrowvert}{\Arrowvert \delta(s) \Arrowvert}sds
\end{equation}
along with its time-average mean value 
\begin{equation}
 \langle Y(t) \rangle=\frac{1}{t}\int_{0}^{t} Y(s) ds.
\end{equation}
The time-weighting factor amplifies any stochastic behaviour, which allows the detection of hyperbolic regions in the time interval ($0,t$). 
$\langle Y(t) \rangle$ enables to distinguish between chaotic and quasi-periodic trajectories: if $\langle Y(t) \rangle \rightarrow \infty$ for $t\rightarrow \infty$ the system is chaotic; while if
$\langle Y(t) \rangle \rightarrow 2$ for $t\rightarrow \infty$ the motion is quasi-periodic.
With this technique we evaluate the upper limits of the eccentricities, and constructed a set of three two-dimensional MEGNO-maps (Supplementary Fig.~\ref{fig:megno-maps}).
We use the MEGNO implementation with the N-body integrator \texttt{REBOUND} \cite{rein2012,reintamayo2015}. The integration time is set to $10^{6}$ times the orbital period of the outermost planet, {\Nplanetd}. The time-step was set as $5\%$ of the period of the innermost planet, {\Nplanetb}, and the simulation was stopped when $\langle Y(t) \rangle \textgreater 10$. 
We run three independent simulations to analyse the upper limits of the eccentricities for pairs of planets, while keeping the third planet's orbit circular in each case. All other planet parameters are fixed to the values in Table~1.%\ref{tab:properties}. 
The size of each MEGNO-map is 100$\times$100 pixels, meaning we explore the eccentricity space for each planet pair up to 10,000 times. 
The results suggest that low eccentricities of 0.05 for all planets are possible. The most restrictive eccentricity is detected for the middle planet {\Nplanetc}, with an upper-limit of 0.05. Planets b and d could reach eccentricities up to 0.1.

In a closely-packed system like {\Ntarget}, tidal interactions between the star and the planets additionally influence the evolution of the orbits. However, the timescale for each parameter differs; for example, the semi-major axis evolves the slowest, while the obliquity and the planetary rotational period can change fast. We explore the tidal evolution using the `constant time-lag model', where the bodies are a weakly viscous fluid \cite{alexander1973}. The mathematical description is given in \cite{mignard1979,hut1981,eggleton1998,leconte2010} and summarised by \cite{bolmont2015}, who implemented it first in their code \texttt{MERCURY-T} and later in \texttt{POSIDONIUS} \cite{blancocuaresma2017}.
We use both codes to verify our findings. {\Nplanetb} likely is Earth-like/rocky, therefore we assume the product of the potential Love number of degree 2 and a time-lag corresponding to Earth's value of $k_{2,\oplus}\Delta \tau_{\oplus} = 213$ $s$ \cite{neron1997}. {\Nplanetc} and {\Nplanetd} likely are rocky/icy planets (taking into account \cite{Fortney2007} and \cite{Chen2017}) with a dissipation higher than Earth's, thus we assume 5$\times k_{2,\oplus}\Delta \tau_{\oplus}$ \cite{bolmont2015,mccarthy2013}. We also assume that the fluid Love number and the potential Love number of degree 2 are equal. 
The rotational period of the host body is uncertain: from photometric and spectral observations we expect an old-slow rotator, but it is possible (yet unlikely) that is a young-fast rotator.
We hence run our simulations for three different rotational periods: P$_{\star,rot}=2,50,100$ days.
First, we explore the evolution of the obliquity and rotational period from different initial conditions: 
initial planetary rotational periods of 10\,h, 100\,h and 1,000\,h, and an initial obliquity of 15$\degree$, 50$\degree$ and 75$\degree$.
The rest of the planet parameters are fixed to the values in Table~1,%\ref{tab:properties}, 
and we assumed eccentricities of 0.05 for all the planets (upper limits from the stability analysis above). 
The results for different stellar rotation periods are comparable. For the slow rotator as an example, we find that the evolution to pseudo-rotational state occurs over a short time-scale of 10$^{4}$-10$^{5}$ yr for all planets, with the outer planet being the slowest to reach this state. 
Since {\Ntarget} is much older than this time-scale, we conclude that our planets are likely well aligned with the host star. 
However, other events which are not studied here, such as magnetic breaks or rotational deformation, might alter this state. 
The resulting rotational periods are P$_{(b,c,d),rot}$=76\,h, 133\,h, and 281\,h, respectively. 

Once the planets reach a pseudo-rotational state, tidal heating keeps acting while the orbits are eccentric, and decreases towards zero with circularisation. 
To explore the circularisation we ran another suite of simulations, performing integrations up to $10^{8}$ yr. We find that after this time the eccentricities shrink by 94--98$\%$ from their initial values, meaning from 0.05 to $<0.002$ for all planets. Since our planetary system is likely much older, this suggests the orbits are in a near-circular configuration. 
While the orbits are still eccentric, the tidal heating is about 250--350\,W\,m$^{-2}$ for planet {\Nplanetb}, 500--600 W m$^{-2}$ for planet c, and 
10\,W\,m$^{-2}$ for planet d. After $10^{8}$\,years, the tidal contribution decreased down to $\sim$1.5\,W\,m$^{-2}$, $\sim$1.0 W m$^{-2}$, and $\sim$0.02\,W\,m$^{-2}$ for planets b, c, and d, respectively.

Finally, we investigate if the {\Ntarget} system remains stable when there is a fourth planet, which is located in the terrestrial-like habitable zone between 0.10--0.28\,AU \cite{Kopparapu2013, Kopparapu2014}.
We again simulate this scenario using MENGO (as described above) for a 5-body system, and a range of orbital distances and masses of the fourth planet (100 values between 0.1-0.3 AU, and 100 values between 1-100 {\mearth}), while freezing all other parameters.
We find that the system is fully stable for the range of masses and semi-major axes in question.

\clearpage
% FIGURES

\begin{figure*}[!htbp]
    \centering
    \includegraphics[width=\textwidth]{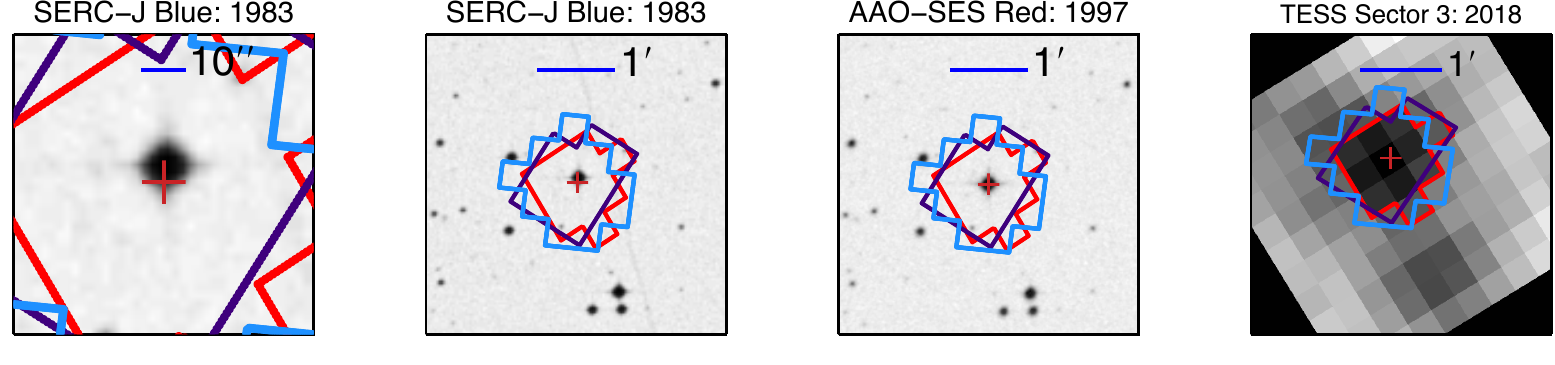}
    \caption{Archival images and {\tess} image for {\Ntarget} from 1983 to 2018.
    The red plus shows the current position of {\Ntarget} in comparison. The regions mark the {\tess} aperture masks used in Sector 3 (red), 4 (purple) and 5 (blue). At the given spatial resolution, we see no background sources at the target's current sky location.}
    \label{fig:archival_images}
\end{figure*}

\begin{figure*}[!htbp]
    \centering
    \includegraphics[width=0.35\textwidth]{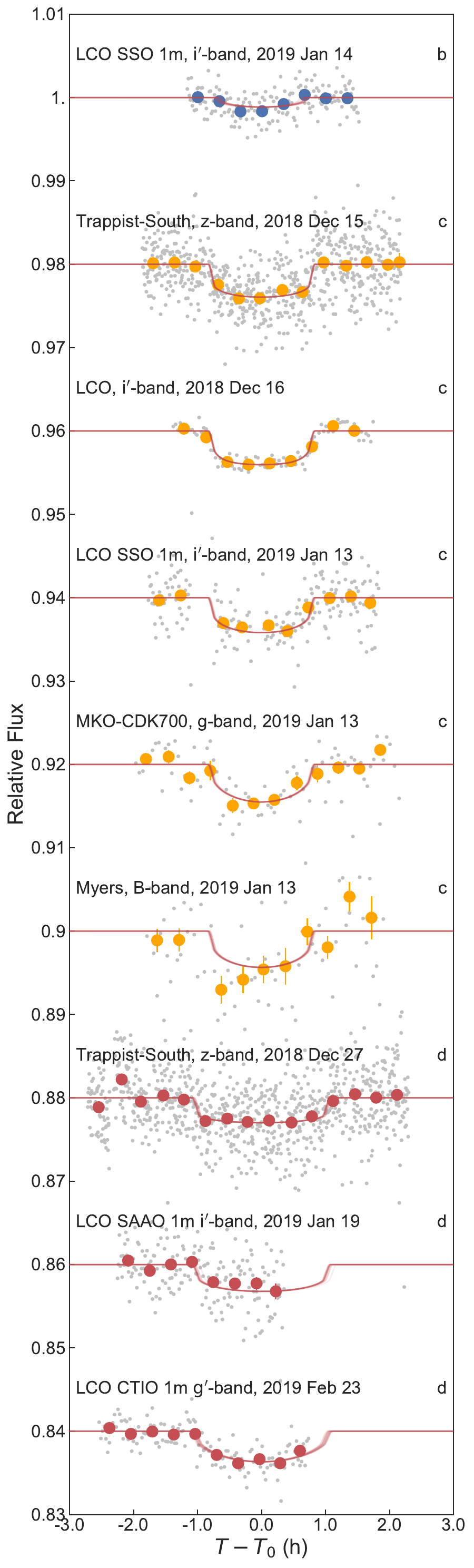}
    \caption{Follow-up lightcurves for {\Ntarget} (see also Supplementary Table~\ref{tab:observations}). Red lines show 20 lightcurves generated from randomly drawn posterior samples from the best-fit \texttt{allesfitter} model.}
    \label{fig:TFOP_lightcurves}
\end{figure*}

\begin{figure*}[!htbp]
    \centering
    \includegraphics[width=0.5\textwidth]{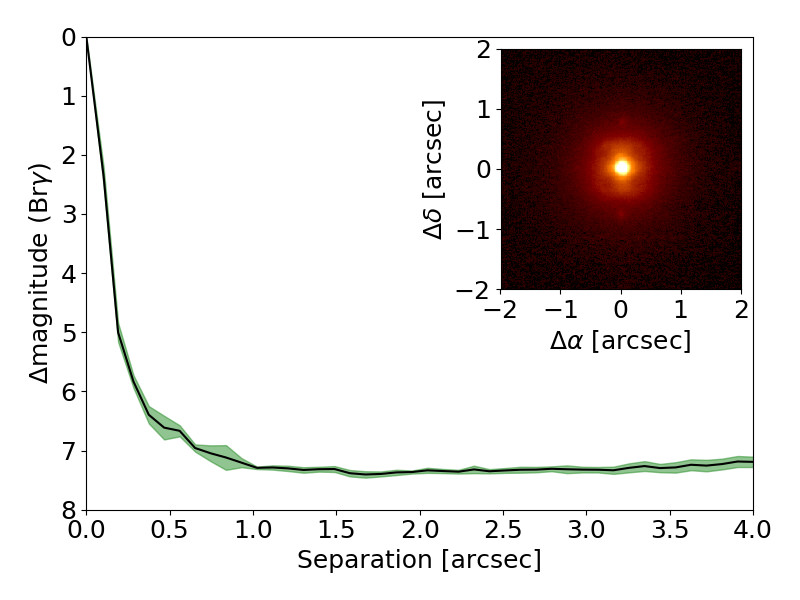}
    \caption{Sensitivity of VLT/NaCo images to nearby companions, as a function of separation. {\it Inset:} 4'' square image, centered on the target. No visual companions appear in this image, or anywhere within the field of view. Note that two point spread function artefacts appear 750\,mas north and south of the host. These artefacts originate from the structure of the point spread function due to the target's brightness, and are not visual companions.}
    \label{fig:naco}
\end{figure*}

\begin{figure*}[!htbp]
    \centering
    \includegraphics[width=1\textwidth]{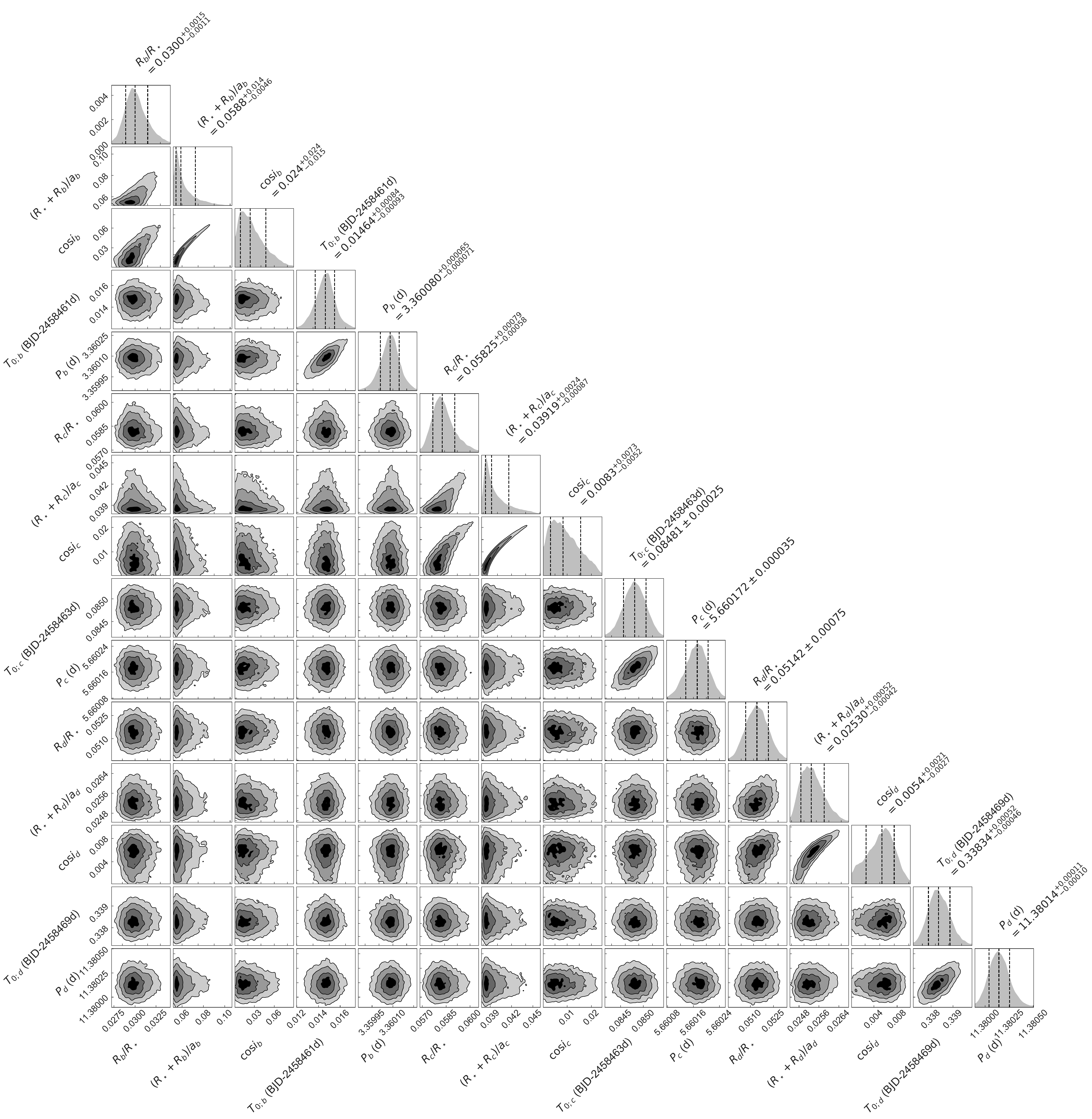}
    \caption{Posterior probability distributions for all astrophysical parameters of the \texttt{allesfitter} nested sampling fit of {\Ntarget}. The figure also highlights the correlation (or absence thereof) between all parameters. Vertical dashed lines show the median and 68\% credible interval.}
    \label{fig:corner}
\end{figure*}

\begin{figure*}[!htbp]
    \centering
    \includegraphics[width=0.5\textwidth]{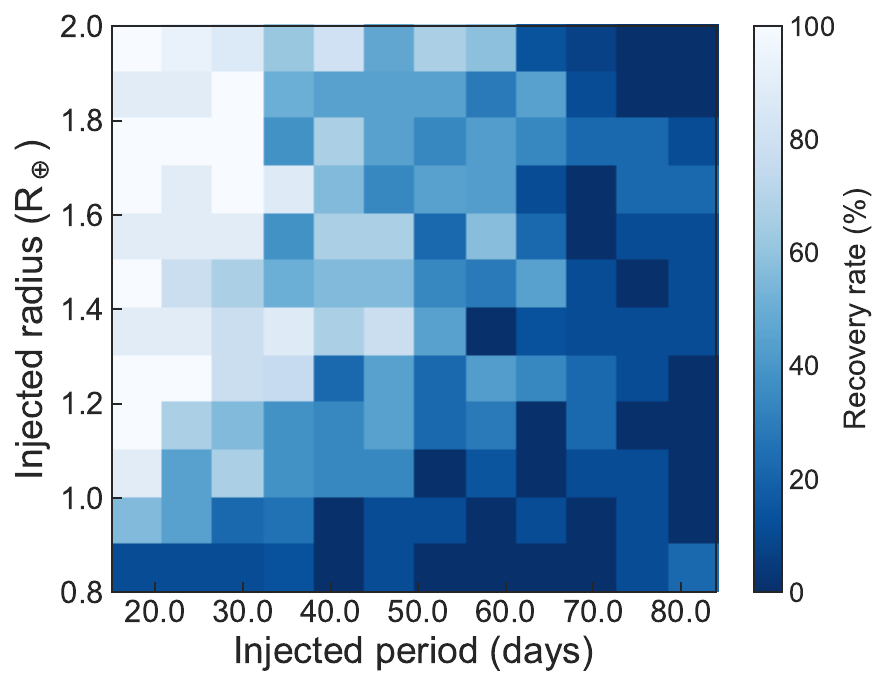}
    \caption{A recovery test for injected transits of small planets in the terrestrial-like habitable zone  of {\Ntarget} (corresponding to periods of 18--85\,days). While larger transiting planets could have been found in the available TESS data, the regime of small exoplanets with period beyond $\sim$30\,days remains open for future transit searches.}
    \label{fig:injected_transit_search}
\end{figure*}

% TABLES
\newgeometry{left=0.5in, right=0.5in} 
\thispagestyle{empty}
\begin{table*}
\tiny
\caption{Observation Log}             
\label{tab:observations}      
\centering 
\begin{tabular}{c l l c c c c c c c c}     % 11 columns 
\hline\hline       
\noalign{\smallskip}
\multicolumn{11}{c}{\it Discovery photometry}\\
\multirow{2}{*}{TOI-270} & Dates & \multirow{2}{*}{Telescope} ${\dag}$ & \multirow{2}{*}{Filter} & Exposure & Nr. of & Duration & Transit & Aperture  & FWHM & \\
& (UTC) &  &  & time (sec) & exposures & (min) & coverage & radius (arcsec) & (arcsec) &  \\
\noalign{\smallskip} 
\hline  
\noalign{\smallskip} 
\multirow{2}{*}{b c d} & 2018-09-20    & \multirow{2}{*}{{\tess}} & \multirow{2}{*}{{\tess}} &  \multirow{2}{*}{120} & \multirow{2}{*}{46874} & \multirow{2}{*}{--} & \multirow{2}{*}{--} & \multirow{2}{*}{30--60''} & \multirow{2}{*}{--} & \\    
                       & -- 2018-12-11 &                          &  & &  & & & & &  \\
\rule{0pt}{2ex} \\
\hline\hline   
\noalign{\smallskip} 
\multicolumn{11}{c}{\it Follow-up photometry}\\
\multirow{2}{*}{TOI-270} & Date & \multirow{2}{*}{Telescope} ${\dag}$ & \multirow{2}{*}{Filter} & Exposure & Nr. of & Duration & Transit & Aperture  & FWHM & \multirow{2}{*}{$\Delta \ln{Z}$}{$^{\S}$}\\
& (UTC) &  &  & time (sec) & exposures & (min) & coverage & radius (arcsec) & (arcsec) &  \\
\noalign{\smallskip} 
\hline  
\noalign{\smallskip}                  
\multirow{8}{*}{b}
& 2018-12-18 & PEST  & Rc & 120 & 189 & 449 & Full & 7.38 & 4.10 & $<0$  \\ 
& 2018-12-25 & LCO-CTIO & i$^{\prime}$ & 20 & 113 & 113 & Ingr.+$66\%$ & 7.78 & 4.30 & $<0$ \\
& 2018-12-27 & SS0-T17  & Clear & 60 & 120 & 151 & Full & 6.30 & 2.10 & N/A${^\ddag}$  \\ 
& 2018-12-28 & PEST  & V & 120 & 174 & 404 & Full & 7.38 & 4.00 & $<0$ \\ 
& 2019-01-11 & LCO-CTIO & i$^{\prime}$ & 14 & 221 & 200 & Full & 8.94 & 2.10 & $<0$ \\ 
& 2019-01-14 & LCO-SSO & i$^{\prime}$  & 15 & 178 & 161 & Full & 7.00 & 1.73 & 6.8{$^{\S}$} \\
& 2019-01-24 & LCO-SSO & g$^{\prime}$ & 40 & 136 & 184 & Full & 4.23 & 2.14 & $<0$ \\
& 2019-01-27 & LCO-SAAO & g$^{\prime}$ & 70 & 119 & 218 & Full & 7.78 & 2.28 & $<0$ \\
\hline
\noalign{\smallskip} 
\multirow{6}{*}{c}
& 2018-12-15 & TS  & z$^{\prime}$ & 10 & 698 & 242 & Full & 4.48 & 2.48 & 11.3{$^{\S}$}  \\ 
& 2018-12-16 & LCO  & i$^{\prime}$ & 90 & 88 & 180 & Full & 5.83 & -- & 19.2{$^{\S}$}  \\
& 2019-01-13 & LCO-SSO  & i$^{\prime}$ & 11 & 207 & 216 & Full & 7.78 & 2.95 & 19.3{$^{\S}$} \\
& 2019-01-13 & PEST  & V & 120 & 143 & 335 & Egr.+$90\%$ & 7.38 & 4.60 & 1.6 \\ 
& 2019-01-13 & MKO  & g$^{\prime}$ & 128 & 82 & 247 & Full & 9.20 & 3.00 & 5.2{$^{\S}$} \\ 
& 2019-01-13 & Myers & B & 180 & 70 & 300 & Full & 4.14 & 4.00 & 7.1{$^{\S}$} \\ 
\hline 
\noalign{\smallskip} 
\multirow{3}{*}{d}
& 2018-12-27 & TS  & z$^{\prime}$ & 10 & 848 & 301 & Full & 4.48 & 2.31 & 4.3{$^{\S}$} \\
& 2019-01-19 & LCO-SAAO  & i$^{\prime}$ & 11 & 182 & 156 & Ingr.+$77\%$ & 5.44 & 1.91 & 6.3{$^{\S}$} \\ 
& 2019-02-23 & LCO-CTIO & g$^{\prime}$ & 70 & 123 & 203& Full & 5.83 & 2.76 & 6.3{$^{\S}$} \\
\rule{0pt}{2ex} \\
\hline\hline
\noalign{\smallskip} 
\multicolumn{11}{c}{\it Reconnaissance spectroscopy }\\
\multirow{2}{*}{TOI-270} & Date & \multirow{2}{*}{Telescope} ${\dag}$ & \multirow{2}{*}{Resolution} & \multirow{2}{*}{Wavelengths} \\
& (UTC) &  &   \\
\noalign{\smallskip} 
\hline  
\noalign{\smallskip} 
   & 2018-12-22 & FIRE & 6000 & $8000-25000\,\text{\AA}$ \\
   & 2019-01-23 & ANU & 23000 & $3900-6700\,\text{\AA}$\\
\rule{0pt}{2ex} \\
\hline\hline
\noalign{\smallskip} 
\multicolumn{11}{c}{\it High-resolution imaging }\\
\multirow{2}{*}{TOI-270} & Date & \multirow{2}{*}{Telescope} ${\dag}$ & \multirow{2}{*}{Filter} & Exposure & Nr. of & FWHM \\
& (UTC) &  &  & time (sec) & exposures & (mas)\\
\noalign{\smallskip} 
\hline  
\noalign{\smallskip} 
-- & 2019-01-25 & NaCo & Ks & 20 & 9 & 90\\
\noalign{\smallskip}
\hline
\end{tabular}
{\dag}{\footnotesize
     Telescopes: \\
     LCO-SSO: Las Cumbres Observatory - Siding Spring (1\,m) \cite{brown2013}\\
     LCO-CTIO: Las Cumbres Observatory - Cerro Tololo Interamerican Observatory (1\,m) \cite{brown2013}\\
     LCO-SAAO: Las Cumbres Observatory - South African Astronomical Observatory (1\,m) \cite{brown2013}\\
     TS: TRAPPIST-South (0.6\,m) \cite{jehin2011}\\
     SSO-T17: Siding Spring Observatory - T17 (0.4\,m)\\
     PEST: The Perth Exoplanet Survey Telescope  (0.3\,m) \\
     Myers: Myers-Siding Spring (0.4\,m) \\
     MKO: Mt. Kent Observatory CDK700 (0.7\,m) \\
     FIRE: Magellan Folded-port InfraRed Echellette (6.5\,m) \cite{Simcoe2013}\\
     ANU: Australia National University Echelle spectrograph (2.3\,m); spectrum reduced following \cite{Zhou2014} \\
     NaCo: VLT NAOS-CONICA (8.2\,m) \cite{Lenzen2003, Rousset2003}\\
     }
{\ddag}{\footnotesize
    Observations not included, as deep exposures were used to study faint neighbouring stars and exclude possible blended eclipsing binaries.\\}
{\S}{\footnotesize
    Only observations with a Bayes factor $\Delta \ln{Z}>3$ (strong evidence for a signal) are used for the global analysis.\\
    }
\end{table*}
\restoregeometry

\clearpage
\begin{table*}[]
    \centering
    \begin{tabular}{lcc}
    \hline
    \hline
    Model                                      & Free parameters  & Bayes factor, $\Delta \ln{Z}$ \\
    \hline
    circular, no TTVs                          & 15                        & --                             \\
    circular, free TTVs for all transits       & 57                        & \textless 0                    \\
    circular, free TTVs for planet b           & 33                        & 2.4                    \\
    circular, free TTVs for planet c           & 30                        & 1.2                    \\
    circular, free TTVs for planet d           & 24                        & \textless 0                    \\
    free eccentricity for all planets, no TTVs & 21                        & \textless 0                    \\
    free eccentricity for planet b, no TTVs    & 17                        & 2.6                    \\
    free eccentricity for planet c, no TTVs    & 17                        & \textless 0                    \\
    free eccentricity for planet d, no TTVs    & 17                        & \textless 0                    \\
    \hline
    \end{tabular}
    \caption{A comparison of various models with different degrees of freedom. The Null Hypothesis, a circular model without TTVs, is compared against more complicated models allowing for free eccentricity and/or free TTVs. A Bayes factor \textgreater 3 would mean strong Bayesian evidence for a model \cite{Kass1995}. We thus find no strong Bayesian evidence for eccentricity nor TTVs.}
    \label{tab:model_comparison}
\end{table*}

\begin{table*}[]
    \centering
    \tiny
    \begin{tabular}{lccccccc}
    \hline
    \hline
    Facility, date & $q_1$ & $q_2$ & $u_1$ & $u_2$ & $\ln{\sigma_\mathrm{white}}$ & GP $\ln{\sigma}$ & GP $\ln{\rho}$  \\
    \hline
    TESS & $0.29_{-0.11}^{+0.20}$ & $0.45_{-0.24}^{+0.26}$ & $0.48_{-0.19}^{+0.16}$ & $0.05_{--0.24}^{+0.34}$ & $-6.6243_{-0.0034}^{+0.0032}$ & $-8.712\pm0.087$ & $-1.97\pm0.25$\\ 
    LCO, 2018-12-16 & $0.54\pm0.28$ & $0.18_{-0.12}^{+0.21}$ & $0.26_{-0.16}^{+0.22}$ & $0.45_{-0.36}^{+0.27}$ & $-7.292\pm0.079$ & $-3.12_{-0.52}^{+0.73}$ & $1.471\pm0.045$ \\
    LCO CTIO, 2019-02-23 & $0.62_{-0.26}^{+0.23}$ & $0.64_{-0.29}^{+0.22}$ & $0.93_{-0.45}^{+0.40}$ & $-0.19\pm0.68$ & $-6.484\pm0.067$ & $-2.06\pm0.55$ & $0.981\pm0.024$ \\
    LCO CTIO, 2018-12-25 & $0.51\pm0.31$ & $0.51\pm0.31$ & $0.64_{-0.41}^{+0.53}$ & $-0.01_{--0.24}^{+0.40}$ & $-6.407_{-0.065}^{+0.070}$ & $-10.5_{-2.8}^{+2.5}$ & $0.3\pm1.5$ \\
    LCO CTIO, 2019-01-11 & $0.44_{-0.27}^{+0.31}$ & $0.48\pm0.30$ & $0.55_{-0.35}^{+0.48}$ & $0.02_{--0.18}^{+0.37}$ & $-6.245\pm0.048$ & $-10.4_{-2.8}^{+3.5}$ & $0.022\pm0.038$ \\
    LCO SAAO, 2019-01-27 & $0.57_{-0.31}^{+0.27}$ & $0.54_{-0.31}^{+0.29}$ & $0.72_{-0.45}^{+0.55}$ & $-0.05_{--0.32}^{+0.42}$ & $-6.472_{-0.064}^{+0.067}$ & $-2.06\pm0.54$ & $1.100\pm0.051$ \\
    LCO SAAO, 2019-01-19 & $0.38_{-0.24}^{+0.31}$ & $0.43_{-0.27}^{+0.31}$ & $0.47_{-0.30}^{+0.40}$ & $0.071_{--0.093}^{+0.37}$ & $-5.833_{-0.051}^{+0.054}$ & $-10.6_{-2.6}^{+2.5}$ & $0.06\pm0.11$ \\
    LCO SSO, 2019-01-24 & $0.44_{-0.28}^{+0.33}$ & $0.46_{-0.29}^{+0.32}$ & $0.52_{-0.35}^{+0.51}$ & $0.05_{--0.13}^{+0.37}$ & $-6.011\pm0.063$ & $-2.19\pm0.55$ & $0.667\pm0.063$ \\
    LCO SSO, 2019-01-13 & $0.45_{-0.25}^{+0.29}$ & $0.40_{-0.25}^{+0.31}$ & $0.49_{-0.31}^{+0.38}$ & $0.111_{--0.074}^{+0.37}$ & $-5.937\pm0.050$ & $-11.0\pm2.5$ & $0.113\pm0.086$ \\
    LCO SSO, 2019-01-14 & $0.62_{-0.33}^{+0.25}$ & $0.55_{-0.33}^{+0.28}$ & $0.77_{-0.49}^{+0.57}$ & $-0.07_{--0.35}^{+0.45}$ & $-6.489\pm0.056$ & $-11.1\pm2.4$ & $0.037\pm0.068$ \\
    MKO-CDK700, 2019-01-13 & $0.59_{-0.31}^{+0.26}$ & $0.53_{-0.30}^{+0.28}$ & $0.74_{-0.44}^{+0.48}$ & $-0.04_{--0.34}^{+0.42}$ & $-6.112_{-0.081}^{+0.086}$ & $-0.46_{-0.61}^{+0.55}$ & $1.983\pm0.020$ \\
    Myers, 2019-01-13 & $0.52\pm0.30$ & $0.48\pm0.31$ & $0.61_{-0.40}^{+0.51}$ & $0.02_{--0.19}^{+0.40}$ & $-5.303_{-0.083}^{+0.088}$ & $-10.2_{-2.9}^{+2.7}$ & $2.00\pm0.10$ \\
    PEST, 2018-12-18 & $0.50\pm0.30$ & $0.50\pm0.30$ & $0.62_{-0.41}^{+0.53}$ & $-0.00_{--0.22}^{+0.40}$ & $-6.208_{-0.057}^{+0.061}$ & $-2.18\pm0.49$ & $0.98\pm0.16$ \\
    PEST, 2018-12-28 & $0.48\pm0.31$ & $0.49\pm0.30$ & $0.59_{-0.39}^{+0.52}$ & $0.01_{--0.20}^{+0.39}$ & $-5.572\pm0.054$ & $-10.7_{-2.6}^{+2.4}$ & $1.11\pm0.15$ \\
    PEST, 2019-01-13 & $0.48_{-0.29}^{+0.30}$ & $0.50\pm0.30$ & $0.60_{-0.38}^{+0.51}$ & $0.01_{--0.21}^{+0.39}$ & $-5.445\pm0.066$ & $-7.8_{-4.3}^{+1.6}$ & $1.36\pm0.10$ \\
    Trappist-South, 2018-12-15 & $0.26_{-0.17}^{+0.30}$ & $0.38_{-0.24}^{+0.32}$ & $0.35_{-0.22}^{+0.30}$ & $0.101_{--0.027}^{+0.32}$ & $-5.827\pm0.027$ & $-4.04\pm0.33$ & $-1.028\pm0.035$ \\
    Trappist-South, 2018-12-27 & $0.17_{-0.13}^{+0.28}$ & $0.38_{-0.25}^{+0.33}$ & $0.28_{-0.19}^{+0.30}$ & $0.074_{--0.015}^{+0.29}$ & $-5.604\pm0.025$ & $-3.82_{-0.63}^{+0.50}$ & $-1.067\pm0.037$ \\
    \hline
    \end{tabular}
    \caption{Nuisance parameters of the fit to individual observations, which are later fixed to their median values in the global analysis. These include the limb darkening parameters $q_1$ and $q_2$ in the parametrization suggested by \cite{Kipping2013} (for comparability also translated into the quadratic limb darkening parameters $u_1$ and $u_2$); the natural logarithm of the white noise scaling $\sigma_\mathrm{white}$; and the hyperparameters of the GP Matern-3/2, namely the natural logarithms of the amplitude $\sigma$ and characteristic time scale $\rho$.}
    \label{tab:nuisance_parameters}
\end{table*}

\begin{table*}[!htbp]
    \centering
    \begin{tabular}{lccccc}
    \hline
    \hline
    TLS\# & SNR & Depth  & Period & First epoch & Note  \\
          &     & (mmag) & (d) & (BJD$_\mathrm{TDB}$) &   \\
    \hline
    1 & 85.8 & 3.9 & 5.65986 & 2458389.50438 & planet c \\
    2 & 54.1 & 3.1 & 11.38025 & 2458389.67737 & planet d \\
    3 & 21.1 & 0.9 & 3.36014 & 2458387.09273 & planet b \\
    4 & 8.3 & 0.2 & 5.53073 & 2458388.19620 & shallow and too wide \\
    5 & 6.5 & 0.6 & 13.90082 & 2458395.07980 & falls in noisy regions$\dagger$ \\
    \hline
    \end{tabular}
    \caption{Threshold crossing events with a signal-to-noise ratio $\mathrm{SNR}\geq5$ detected with \texttt{transit least squares} \cite{Hippke2019} in TESS Sectors~3--4 short-cadence data. The search is performed on the PDC-SAP lightcurves, which were additionally detrended using a Gaussian process. $\dagger$ Note that this signal might arise from systematics related to the satellite orbit ($\sim$13.7 days).}
    \label{tab:TLS}
\end{table*}

\clearpage

%%%%% REFERENCES %%%%%

% working version, for preparing the final draft:

% \bibliographystyle{naturemag}
% \bibliography{references}

% for final submission, click "submit --> to arxiv --> download zip file, copy the contents of .bbl into here, and comment out the above lines:

 \renewcommand{\noop}[1]{}